\documentclass[%
 reprint,
 nofootinbib,
 amsmath,amssymb,
 aps,
]{revtex4-2}

\usepackage{graphicx}
\usepackage{dcolumn}
\usepackage{bm}
\usepackage{amsmath}
\usepackage{xcolor}

\begin{document}

\title{Quasi Normal Modes in Dispersive Photonic Time-Crystals}

\author{Calvin M. Hooper}
\email{ch1122@exeter.ac.uk}
\author{Ian R. Hooper}
\author{Simon A.R. Horsley}
\affiliation{
 School of Physics and Astronomy, University of Exeter, Stocker Road, Exeter, EX4 4QL, UK
}


\begin{abstract}
Quasinormal modes characterise the transient response of static optical cavities. Here, we introduce the notion of a Floquet quasinormal mode to describe transient responses in photonic time crystals. Contrasting their static counterparts, exceptional points associated with symmetry transitions are an inherent feature, as modes spontaneously and non-perturbatively lock their phase to the oscillations of the material. We further investigate the limiting behaviour of the Floquet quasinormal modes in large cavities. New non-perturbative behaviour arises in time-modulated systems as increasingly large time-crystal cavities come closer to achieving the maximum gain predicted from a bulk wavenumber bandgap.
\end{abstract}

\maketitle

%
%
\section{\label{sec:introduction}Introduction}

Noether’s theorem~\cite{Noether1918} links energy conservation to the translational invariance of a system in time.  This applies as much to particle physics as to classical wave propagation.  Yet when a wave--supporting medium is explicitly time dependent, time translation symmetry is broken.  The propagation characteristics of the wave then become sensitive to the time delay relative to the material modulation, a sensitivity that is connected to the pumping of energy into or out of the system~\cite{Galiffi2022, Caloz2020, Cartella2018, Castaldi2023, Zhu2023, Serra2024, Pendry2024, Dong2025, Lustig2018, Harwood2025, Asgari2024}. We should remember, however, that such energy must always be supplied externally to the medium: whilst the wave inside a material appears to receive free energy, the lab still pays a power bill.

Perhaps the clearest example of this unusual behaviour occurs in photonic time-crystals~\cite{Galiffi2022, Lustig2023, He2023, Pendry2024, GaxiolaLuna2023, Lustig2018, Asgari2024}. What makes these periodically driven media so striking, aside from their neat analytical properties, is the exponential growth of waves within the medium as a function of time.  Indeed, as long as the wave frequency is close to commensurate with the material’s time-variations, pumping of energy to/from the wave will compound exponentially in time~\cite{Galiffi2022, Dong2025, Trainiti2019, He2023, Serra2024, Pendry2024, GaxiolaLuna2023, Lustig2018, Asgari2024}.  Whether energy is pumped into or out of an incident wave is determined entirely by the wave’s relative phase with respect to the oscillation of the medium~\cite{Hooper2025, Galiffi2024, Hendry2025, Wang2023}.

Analysis of periodically modulated media has largely followed via an analogy with spatial crystals~\cite{Lustig2023, Trainiti2019, GaxiolaLuna2023, Asgari2024}.  For instance, the preceding discussion of exponential gain in periodically driven media can be understood directly in terms of the theory of band gaps.  In spatial crystals, the discrete rather than continuous translational symmetry replaces momentum as a conserved quantity with the quasimomentum, which is periodic in reciprocal space. By direct analogy, frequency is no longer conserved in time-crystals, replaced with the periodic quasifrequency.

Similarly, whilst spatial crystals possess a \emph{frequency} bandgap within which the wavenumber becomes complex, temporal crystals possess a \emph{wavenumber} bandgap within which the frequency becomes complex (corresponding to the aforementioned exponential growth/decay of fields as a function of time). Furthermore, in both cases, the maximum growth and decay rates within a spatial/temporal crystal bandgap may be found analogously, by substituting in quasimomenta/quasifrequencies with an increasing range of imaginary parts until the corresponding frequency/wavenumber ceases to be real. In each case, this occurs at an exceptional point as the real frequencies/wavenumbers within the bandgap break into complex conjugate pairs for complex quasimomenta/quasifrequencies with too much growth/decay to be supported by the spatial/temporal crystal.

Perhaps more important than the similarities between space and time-crystals are their differences. For instance, the preservation of frequency by spatial crystals allows second quantisation to proceed in a straightforward manner. By contrast, time-crystals inherently couple positive and negative wave frequencies, thus mixing the associated creation and annihilation operators.  This mixing allows photons to be extracted not only from thermal fluctuations~\cite{VertizConde2025, VazquezLozano2023}, but also from the vacuum state~\cite{Horsley2023BlackHole, Horsley2024, VazquezLozano2023}.

As soon as we consider wave propagation in the time domain, we must face the linked complexities of dispersion and dissipation~\cite{Horsley2023Operators}.  Both effects modify the outcome of proposed experiments: dispersion naturally limits the compression of pulses by analogue black holes, as predicted by Horsley et al.~\cite{Horsley2023BlackHole, Horsley2024}, whilst intrinsic loss, inherent to certain methods of time-modulation~\cite{Tomadin2018, Moussa2023}, can overwhelm any anticipated wave amplification, closing wavenumber bandgaps~\cite{Hooper2025, Wang2025}.  Previously, we investigated the amplification of waves in a dispersive, dissipative slab through calculating the continuous wave transmission operator~\cite{Hooper2025}.  In this paper we study the same system through extending the quasi normal mode concept to periodically driven materials (Floquet Quasi Normal Modes, FQNMs).  

Normal modes of a static, closed system are those fields that maintain their profile and oscillate harmonically for---in the ideal case---infinite time.  In real experiments all modes have a finite lifetime and accordingly the normal mode frequencies of practical systems are always complex valued, the imaginary part corresponding to the inverse decay time of the mode.  Such complex frequency normal modes are dubbed Quasi Normal Modes (QNMs~\cite{Benisty2022, Kristensen2014}), and they are distinguished from true normal modes due to their modified normalization, orthogonality, and completeness relations.  They are useful because a discrete set of such modes can be employed to understand the otherwise complex dynamics of open systems.

In time--varying media, however, the wave frequency is not conserved.  Is it thus even meaningful to extend the normal mode concept to time--varying materials?  It seems it is: as discussed in~\cite{Horsley2023Operators}, propagation in the material can still be described using a linear operator, which has associated eigenfunctions.  These eigenfunctions contain a spread of frequencies that are interconverted such that the spectrum is unmodified after a single modulation period.  Given the difficulty of understanding wave propagation in dispersive, time--varying materials, might it be useful to thus extend the QNM concept to time--varying media (FQNMs)? 

Here we outline such an extension, quantifying the trajectories traced in the complex plane by the FQNM quasifrequencies under continuous variations of a cavity.   These modes no longer conserve complex frequency, but rather the more general complex Floquet quasifrequency.  We identify a number of regions where non-perturbative behaviour is in fact fundamental to understanding the behaviour of FQNMs. Indeed, this behaviour necessarily limits the application of conventional perturbation theory to the FQNM problem.  Our results are thus complementary to the perturbative analysis of~\cite{Valero2025}, and the scattering theory of~\cite{Vial2025}.

The structure of this paper is as follows: In Section~\ref{sec:FloquetQNMs} we apply the operator formalism of Horsley et al.~\cite{Horsley2023Operators} to show that the FQNM problem can be written in a particularly compact and intuitive form.  We then consider a number of general properties regarding the trajectories of FQNM quasifrequencies in the complex plane  (our justification for these trajectories being well-defined in the first place is provided in Appendix~\ref{app:FQNMFundamentals}).  In Section~\ref{sec:ExceptionalPoints} we prove how, in contrast to static media, exceptional points are a ubiquitous feature in the trajectories of FQNMs under continuous variations of the system parameters.  We show that these exceptional points are a necessary consequence of the symmetry presented in~\cite{Hooper2025}.  In Section~\ref{sec:LargeCavities} we consider the global distribution of FQNMs in the complex plane through the limit of large slab lengths. We demonstrate that the introduction of time-variations has a radical effect on this limit, again due to the symmetry described in~\cite{Hooper2025}. Indeed, these limits directly describe how the gain (and loss) present in the finite time-crystal cavities of experiment eventually approaches the theoretical predictions for an infinite bulk medium. Finally, in Section~\ref{sec:Results}, we verify our results in a particular example, characterising the origin of the transmission poles observed in~\cite{Hooper2025}, in terms of their associated modes.

%
%
\section{\label{sec:FloquetQNMs}Quasi Normal Modes in a Time--Varying Material}

In static media, an efficient method for computing the time-evolution of waves in a cavity is to study its QNMs. These modes are solutions to the undriven wave equation, assuming time harmonic evolution with frequency $\omega$.  For a perfectly closed, dissipation free system these frequencies are purely real valued and correspond to the usual normal mode frequencies.  Meanwhile for a general system that is both open (outgoing boundary conditions) and dissipative, the \emph{quasi}--normal mode frequencies \(\omega\) are complex, where \({\rm Im}[\omega]<0\)(/\({\rm Im}[\omega]>0\)) corresponds to loss(/gain).  These QNMs provide a framework for analysing the linear response of a cavity, including both short--time transients and the long--time steady state\footnote{To state this for \emph{any} initial condition of the cavity requires the notion of quasinormal mode completeness, which has not been rigorously proven true in general~\cite{Sauvan2022}.}.

Due to the assumption of harmonic time evolution, the QNM eigenvalue problem requires a system with time translational symmetry, an assumption that fails for time-varying media.  But under the constraint of periodicity in time, the Floquet quasifrequency, $\omega_0$~\cite{Lyubarov2022} takes its place.  Hereafter \(T\) denotes the period of modulation of the material parameters, with \(\Omega = \frac{2\pi}{T}\) its angular frequency.  The harmonic ansatz for FQNMs is then given by
\begin{equation}
    \Psi\left( \mathbf{r},t;\omega_{0} \right) = {\rm e}^{- \mathrm{i}\omega_{0}t}\overline{\Psi}\left( \mathbf{r},t;\omega_{0} \right)
    \label{eq:FloquetAnsatz},
\end{equation}
where $\omega_0$ is the Bloch frequency and \(\Psi\) is a vector containing both the electric and magnetic fields, as well as the material response (e.g. the electric polarization and its time derivative). \(\overline{\Psi}\) differs from \(\Psi\) as it is periodic with period $T$, i.e. \(\overline{\Psi}\left( \mathbf{r},t;\omega_{0} \right) = \overline{\Psi}\left( \mathbf{r},t + T;\omega_{0} \right)\). Eq. (\ref{eq:FloquetAnsatz}) thus corresponds to fields that are restored up to a scalar multiple of \({\rm e}^{-\mathrm{i}\omega_0 T}\) after a single period has elapsed.  Note that there is the usual non-uniqueness of $\omega_0$ in the definition (\ref{eq:FloquetAnsatz}), where we may shift its value by any multiple of $\Omega$, keeping the function $\bar{\Psi}$ periodic: \({\rm e}^{- \mathrm{i}\omega_{0}t}\overline{\Psi}\left( \mathbf{r},t;\omega_{0} \right) = {\rm e}^{- \mathrm{i}\left( \omega_{0} + m\Omega \right)t}\left( {\rm e}^{\mathrm{i}m\Omega t}\overline{\Psi}\left( \mathbf{r},t;\omega_{0} \right) \right)\).

In Appendix \ref{app:FQNMEigenvalueProblem}, we describe how such an FQNM eigenvalue problem may be set up in general. However, for the contents of this paper, we will direct our attention towards wave propagation at normal incidence (along the $z$ axis), propagating through a temporally periodic dielectric slab.  Here the transverse electric field $E$ obeys a generalization of the one dimensional wave equation,

\begin{equation}
    \frac{\partial^{2}E}{\partial z^{2}} = \frac{1}{c^{2}}\frac{\partial^{2}}{\partial t^{2}}\left[\left( 1 + \Pi_{L}(z)\chi\left( \partial_{t},t \right) \right)E\right],
    \label{eq:TimeDomain1DWaveEquation}
\end{equation}
where the function $\Pi_{\rm L}$ is zero everywhere except for inside the dielectric
\begin{equation}
    \Pi_{L}(z)=\begin{cases}
    1&|z|<\frac{L}{2}\\
    0&|z|\geq\frac{L}{2}
    \end{cases}.
\end{equation}
The quantity $\chi(\partial_{t},t)$ appearing in Eq. (\ref{eq:TimeDomain1DWaveEquation}) is the operator representing the dielectric susceptibility of the dispersive, time--varying medium. Although our findings are general, we will---where necessary---consider the same example as in~\cite{Hooper2025}, and assume the time-varying Drude susceptibility,
\begin{equation}
    \chi\left( \partial_{t},t \right) = \left( 1 + \eta\cos(\Omega t) \right)\frac{\omega_{\mathrm{pl},0}^{2}}{\partial_{t}^{2} + \gamma\partial_{t}}.
    \label{eq:ExampleDrudeModel}
\end{equation}
\noindent with the parameters \(\eta = 0.2\), \(\gamma = 0.01\ \mathrm{s}^{- 1}\), \(\omega_{\rm pl,0} = 0.3\ \mathrm{rad}\ \mathrm{s}^{- 1}\), and \(c = 1\ \mathrm{m}\ \mathrm{s}^{- 1}\).  Note that we've chosen a particular ordering of the $\partial_{t}$ and $t$ operators in (\ref{eq:ExampleDrudeModel}), which corresponds to a particular microscopic model of the material dynamics.  We note this ordering allows for the amplification of an incident wave, something which is not guaranteed~\cite{Moussa2023}.

We now follow~\cite{Horsley2023Operators}, calculating the effect of wave propagation on the frequency spectrum of the electric field. Following Eq. (\ref{eq:FloquetAnsatz}), individual solutions for \(E\) are given by

\begin{equation}
    E\left( z,t;\omega_{0} \right) = {\rm e}^{- \mathrm{i}\omega_{0}t}\sum_{n = - \infty}^{\infty}{{\widetilde{E}}_{n}\left( z\right){\rm e}^{- n\mathrm{i}\Omega t}}.
    \label{eq:FloquetAnsatzInComponentForm}
\end{equation}
An important symmetry in this representation was presented in~\cite{Hooper2025}, which we now generalise to the complex quasifrequencies of QNMs.

Time-domain fields are real valued and are hence symmetric under \(\mathcal{C}\)-symmetry, where \(\mathcal{C}\) is the complex conjugate operator \(\mathcal{C}a=a^*\mathcal{C}\). However, the Floquet ansatz (\ref{eq:FloquetAnsatz}) immediately breaks this symmetry for any \(\omega_0\) where \({\rm e}^{-\mathrm{i}\omega_0T}\) is not real valued.  To obtain a real time-domain field in this case, one must thus add together modes (\ref{eq:FloquetAnsatzInComponentForm}) with Floquet quasifrequencies \(\omega_0\) and \(-\omega_0^*\).

Yet the modes given in Eq. (\ref{eq:FloquetAnsatzInComponentForm}) do not always break complex conjugation symmetry. If, for example, \(\Re\left\{\omega_0\right\}=0\), the functions \(E\left( z,t;\omega_{0} \right)\) are real valued when the Fourier components obey \({\widetilde{E}}_{n}=\mathcal{RC}{\widetilde{E}}_{n}\mathcal{RC}\), where \(\mathcal{R}\) reverses the Fourier spectrum of a wave, i.e. \(\mathcal{R}{\widetilde{E}}_{n}\mathcal{R}={\widetilde{E}}_{-n}\).  If this is the case we say the mode is $\mathcal{R}\mathcal{C}$ symmetric, a term introduced in Ref.~\cite{Hooper2025}, and which indicates a standing wave dependence of the wave in time, analogous to the standing wave solutions at the edge of the Brillouin zone in a spatially periodic medium. 

Similar to wave solutions in a spatially periodic medium, as we can translate the Bloch frequency $\omega_0$ by any multiple of $\Omega$, this $\mathcal{RC}$ symmetry can hold about any axis parallel to the imaginary axis where the real part of $\omega_0$ is an integer multiple of $\Omega/2$.  As a result, for these special choices of $\omega_0$ the solution (\ref{eq:FloquetAnsatz}) can be chosen to correspond to a real time-domain field, whilst for other values of the Bloch frequency, two different solutions must always be combined. Henceforth, we denote with $A_n$ the axis corresponding to frequencies with a real part $\frac{n\Omega}{2}$.

We now treat the Fourier components \({\widetilde{E}}_{n}\) defined in Eq. (\ref{eq:FloquetAnsatz}) as the components of a single infinite vector \(|\widetilde{E}\rangle\). The latter is then governed by Eq. (\ref{eq:TimeDomain1DWaveEquation}), with operators substituted by their frequency domain counterparts

\begin{equation}
    \begin{split}
        E &\mapsto |\widetilde{E}\rangle, \\
        \partial_{t} &\mapsto - \mathrm{i}\widehat{\omega}, \\
        \cos(n\Omega t + \phi) &\mapsto {\widehat{\Delta}}_{n}(\phi),
    \end{split}
    \label{eq:FloquetOperatorConversions}
\end{equation}

\noindent where
\begin{equation}
    \begin{matrix}
        \widehat{\omega} = \ {\mathrm{diag}}
        \begin{pmatrix}
            \cdots & \omega_{0} - \Omega & \omega_{0} & \omega_{0} + \Omega & \cdots
        \end{pmatrix}, \\
        \left( {\widehat{\Delta}}_{n}(\phi) \right)_{ij} = \frac{1}{2}\left\{
        \begin{matrix}
            {\rm e}^{\mathrm{i}\phi} & i - j = - n \\
            {\rm e}^{- \mathrm{i}\phi} & i - j = n \\
            0 & \mathrm{otherwise}
        \end{matrix} \right.\ ,
    \end{matrix}
    \label{eq:FloquetOperatorDefinitions}
\end{equation}

\noindent for \(n\geq 1\).  Note that we include the index `$n$' in Eq. (\ref{eq:FloquetOperatorConversions}) so that e.g. Eq. (\ref{eq:ExampleDrudeModel}) may be extended to an arbitrary time dependence of the plasma frequency, written as a Fourier sum.  Combining Eqns. (\ref{eq:TimeDomain1DWaveEquation}) and (\ref{eq:FloquetOperatorConversions}) leads to an operator valued version of the Helmholtz equation,
\begin{equation}
    \frac{\partial|\widetilde{E}\rangle}{\partial z^2}+\frac{\widehat{\omega}^2}{c^2}\left[1+\Pi_{L}(z)\left(1+\eta\widehat{\Delta}_1(0)\right)\frac{\omega_{{\rm pl},0}^2}{\widehat{\omega}^2+{\rm i}\gamma\widehat{\omega}}\right]|\widetilde{E}\rangle=0.\label{eq:operator_helmholtz}
\end{equation}

As in these definitions, throughout the rest of this paper we will often leave the dependence of any Floquet operator on \(\omega_{0}\) as implicit for the sake of brevity.

At this point, we note that any time-domain operator which maps a real valued physical field to another real valued field must preserve $\mathcal{RC}$ symmetry. Thus, for any such operator \(\mathcal{O}\left(\omega_0\right)\), we require \(\mathcal{RC}\mathcal{O}\left(\omega_0\right)\mathcal{RC}=\mathcal{O}\left(-\omega_0^*\right)\), which, for the special case of \(\omega_0\) lying on any symmetry axis $A_n$ reduces to
\begin{equation}
    \mathcal{R}_n\mathcal{C}\ \mathcal{O}\left( \omega_{0} \right)\mathcal{R}_n\mathcal{C}=\ \mathcal{O}\left( \omega_{0} \right),
    \label{eq:RCOperator}
\end{equation}
where \(\mathcal{R}_n\) denotes a reflection of the components of $\tilde{E}$ about the axis \(A_n\). This symmetry can be easily verified for the example operators, \(-\mathrm{i}\widehat{\omega}\) and \(\widehat{\Delta}_n\left(\phi\right)\) defined in Eq. (\ref{eq:FloquetOperatorDefinitions}). This symmetry of operators under transformations involving a pair of operators is closely related to the theory of \(\mathcal{PT}\)-symmetry~\cite{Bender1998}.

Solving Eq. (\ref{eq:TimeDomain1DWaveEquation}) proceeds as usual (as described in~\cite{Horsley2023Operators}), albeit with care regarding operator ordering.  We substitute the operator valued generalization of a sum of travelling waves, \(|\widetilde{E}\rangle = {\rm e}^{\mathrm{i}\widehat{K}z}|{\widetilde{E}}_{+}\rangle + {\rm e}^{- \mathrm{i}\widehat{K}z}|{\widetilde{E}}_{-}\rangle\) for the region within the slab \(|z| < \frac{L}{2}\), and the corresponding outgoing waves, \(|\widetilde{E}\rangle = {\rm e}^{\mathrm{i}\frac{\widehat{\omega}}{c}\left( \pm z - \frac{L}{2} \right)}|{\widetilde{E}}_{\mathrm{out}, \pm}\rangle\), in the region outside \(|z| > \frac{L}{2}\).  Requiring continuity of these expression at the slab boundary, solutions to Eq. (\ref{eq:operator_helmholtz}) reduce to the requirement that the outgoing field vector, $\tilde{E}_{\rm out}$ is in the nullspace of $\widehat{\mathrm{Q}}$,
\begin{equation}
    \widehat{\mathrm{Q}}|\tilde{E}_{\mathrm{out}, +}\rangle = 0,
    \label{eq:FQNMCondition}
\end{equation}
\noindent where the operator, $\widehat{\rm Q}$ (which equals the inverse of the transmission operator \(\widehat{t}\) given in Ref. \cite{Hooper2025}) is given by
\begin{equation}
        \widehat{\mathrm{Q}} = \left( \frac{1 + \widehat{n}}{2}{\rm e}^{- \mathrm{i}\widehat{K}L}\frac{1 + \widehat{n}}{2} - \frac{1 - \widehat{n}}{2}{\rm e}^{\mathrm{i}\widehat{K}L}\frac{1 - \widehat{n}}{2} \right){\widehat{n}}^{- 1},\label{eq:Q-kernel}
\end{equation}
where we have introduced the following operator definitions of the frequency domain refractive index, wavenumber squared, and susceptibility, 
\begin{align}
    \widehat{n} &= \left( \frac{\widehat{\omega}}{c} \right)^{- 1}\widehat{K}, \nonumber\\
    {\widehat{K}}^{2} &= \frac{{\widehat{\omega}}^{2}}{c^{2}}\left( 1 + \widehat{\chi} \right), \nonumber\\
    \widehat{\chi}&=\left( 1 + \eta{\widehat{\Delta}}_{1}(0) \right)\frac{\omega_{\rm pl}^{2}}{{\widehat{\omega}}^{2} + \mathrm{i}\gamma\widehat{\omega}}.\label{eq:FQNMOperatorDefinitions}
\end{align}

\begin{figure}
    \centering
    \includegraphics[width=0.9\linewidth]{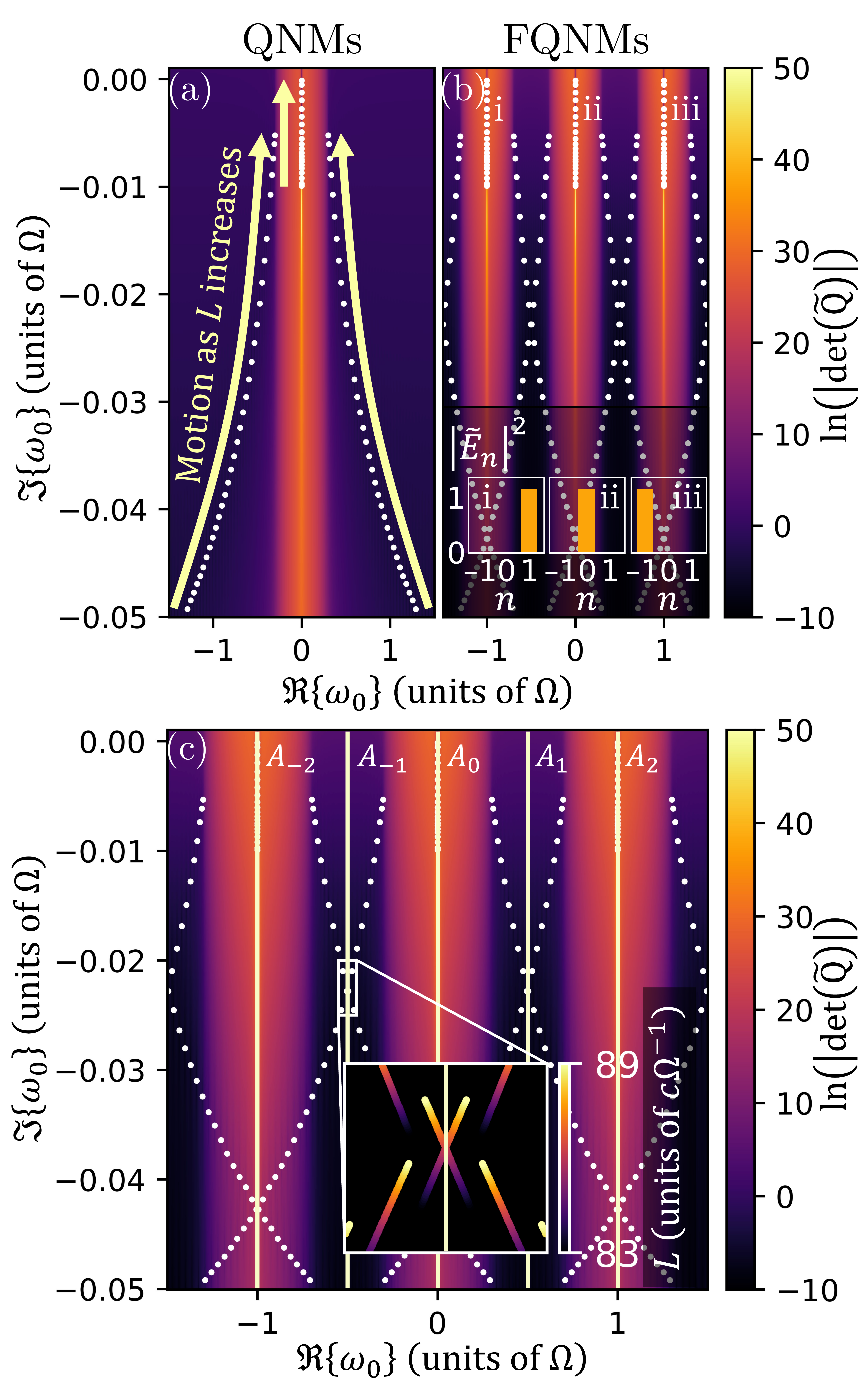}
    \caption{
        \textbf{The QNMs and FQNMs of a static Drude metal slab:}
        The parameters are those given in the main text, with \(\eta = 0\), setting the time modulation to zero, and a slab length of \(L = 85\ c\Omega^{- 1}\) (aside from the inset of panel c).
        In all panels we show a colour plot of the indicator function \(\ln( | \det( \widetilde{\mathrm{Q}}) |)\), the divergence of this quantity indicating a non--empty kernel, and thus a solution to (\ref{eq:FQNMCondition}). Numerically determined roots of \(\det( \widetilde{\mathrm{Q}})=0\) are plotted with white dots. Here \(\widetilde{\mathrm{Q}}\) approximates the operator \(\mathrm{\widehat{Q}}\) derived in the main text, but truncated to a \(5 \times 5\) matrix.
        (a) The QNMs of a static Drude metal. The band of large \(| \det( \widetilde{\mathrm{Q}})|\) around \(\Re(\omega_0)=0\) corresponds to a region of suppressed transmission. The modes within this region are thus mostly confined to the material, and only weakly couple to electromagnetic waves.
        (b) The FQNMs of a static Drude metal. These can be immediately identified as shifted copies of the QNMs of panel (a), shifted in frequency by integer multiples of \(\Omega\). Denoted with i-iii are \(3\) such copies, with their associated eigenvectors plotted component-wise in the inset. Note that the frequency shifts of each eigenvector precisely cancel the shifted quasifrequency $\omega_0$, such that each replica FQNM corresponds to the same time-domain field.
        (c) The crossing of FQNM quasifrequencies about symmetry axes (labeled with \(A_{n}\)). Inset: FQNMs crossing a symmetry axis. The FQNM quasifrequencies are plotted as coloured points denoting slab lengths between \(L=83\ c\Omega^{- 1}\) and \(L=89\ c\Omega^{- 1}\). By definition of the symmetry axis, as a given mode crosses an \(\mathcal{RC}\)-axis \(A_{n}\), it must collide with its \(\mathcal{RC}\)-symmetric pair. In static media, these modes aren't coupled, and so no additional behaviour arises at these crossings, a fact expected to change for finite modulations.
    }
    \label{fig:StaticFQNMs}
\end{figure}

Eq. (\ref{eq:FQNMCondition}) is solved by varying the complex value of the Bloch frequency $\omega_0$ until there is at least one zero eigenvalue of $\rm \widehat{Q}$.  The associated vector $|\tilde{E}\rangle$ then tells us the Fourier spectrum of the FQNM of our time--varying slab.  In Fig.~\ref{fig:StaticFQNMs} we present the simplest possible example using this formalism, generalising the QNMs (panel a) of a static system to their FQNM equivalents (panel b) without adding any time-variation. In this case, the modes of the static system are copied into a series of replicas separated by gaps of the modulation frequency. This is analogous to the copies of the dispersion relation of a homogeneous medium obtained through assuming spatial periodicity, with the separation given by the reciprocal lattice vector.  Each replica is a different representation of the same physical field, with offsets of \(n\Omega\) in the complex plane precisely balanced by discrete shifts in the Floquet spectrum. In addition to the Floquet symmetry demonstrated in this figure, the predicted symmetry axes \(A_{n}\) are also immediately observed, and clearly remain as slab thickness is varied.

Aside from edge cases where a material response drops to zero at certain frequencies, $\widehat{\rm Q}$ will generally be infinite-dimensional. Thus, for practical reasons, in calculations we will often refer to $\widetilde{\rm Q}$, defined as a finite-dimensional approximation to $\widehat{\rm Q}$ where the susceptibility operator $\widehat{\chi}$ is assumed to be finite-dimensional. We discuss this approximation in more detail in~\cite{Hooper2025}. For intuitive purposes, in this paper, it suffices to note that the dimension of $\widetilde{\rm Q}$ corresponds directly to the number of replicas present for any given mode.

There are a few subtleties within the definitions (\ref{eq:FQNMOperatorDefinitions}). Firstly, we might worry that the infinite number of choices of operator square root, \(\widehat{K} = [\widehat{K}^{2}]^{1/2}\) each leads to a different prediction when applying Eq. (\ref{eq:FQNMCondition}) .  However, close examination shows that \(\widehat{\mathrm{Q}}\) is a holomorphic function of the operator \({\widehat{K}}^{2}\), and thus independent of this choice of square root. Similarly, divergences of \(\widehat{n}^{-1}\) cancel when considering the full expression for \(\widehat{\mathrm{Q}}\).

By contrast, the behaviour of the operator \(\widehat{\mathrm{Q}}\) when \(\widehat{K}^2\) diverges, e.g. at a complex resonant frequency, is more problematic.  In the usual QNM case, this corresponds to an essential singularity, with an infinite number of modes appearing at nearby complex frequencies. In Appendix \ref{app:FQNMFundamentals} we apply Fredholm theory~\cite{Renardy2004} to demonstrate that this idea generalises to the case of FQNMs in dispersive \(1\)-dimensional slabs.

For future reference, we define the set \(S_{\mathrm{R}}\) as FQNM quasifrequencies of the system when with coupling is removed between the waves and material. In Appendix~\ref{app:FQNMFundamentals}, we demonstrate that, away from these points all FQNMs have a finite spacing, and move continuously under perturbation. Thus, for our remaining analyses, we focus on understanding FQNMs in the remainder of the complex plane (which we term the set \(S_0\)).

%
%
\section{\label{sec:ExceptionalPoints}Degeneracies and Exceptional Points}

In a static system, \(\mathcal{RC}\)-symmetry implies whenever a mode crosses a symmetry-axis \(A_n\), it must collide with another such mode (see Figure~\ref{fig:StaticFQNMs}.c). This close proximity ensures that such modes are greatly affected by the frequency-coupling introduced by time-modulation.

\begin{figure}
    \centering
    \includegraphics[width=0.9\linewidth]{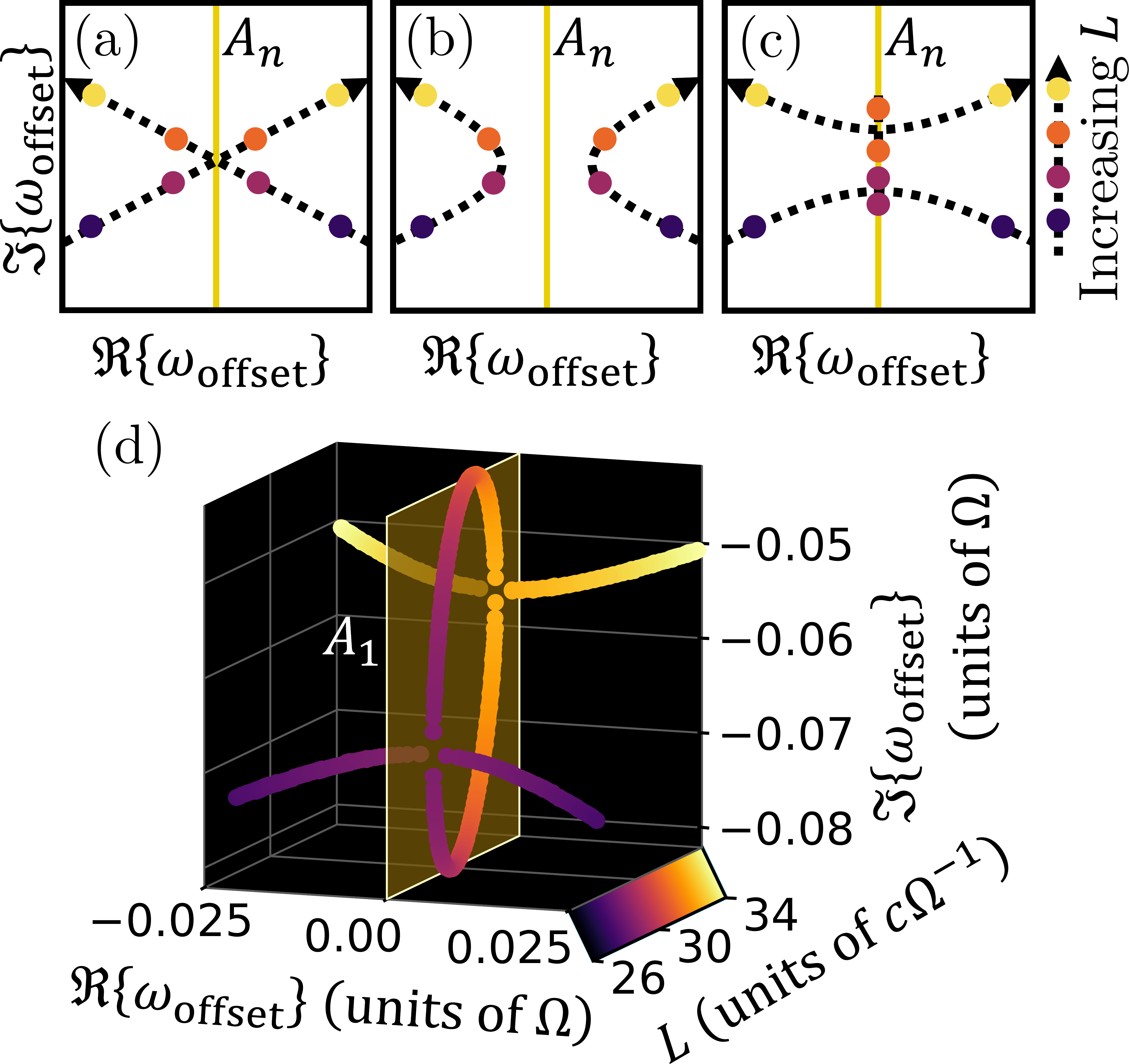}
    \caption{
        \textbf{\(\mathcal{RC}\)-symmetry (un)breaking of FQNMs in a photonic time crystal:}
        In all plots, we consider frequency offsets \(\omega_{\rm offset}\) relative to an \(\mathcal{RC}\)-symmetry axis \(A_{n}\) (specified as \(A_1\) in panel d). In panels a-c we plot the possible qualitative behaviours allowed by symmetry for FQNMs approaching a symmetry axis \(A_{n}\). (a) If the modes are not coupled by time-variations in the system, they will pass one another unperturbed. This is most likely for static media (see Figure \ref{fig:StaticFQNMs}). Away from the crossing point, the behaviour of later modes can be expected to line up with this example. (b) As is commonly expected from Hermitian systems, a pair of modes may experience avoided crossing. (c) Our system obeys \(\mathcal{RC}\)-symmetry rather than Hermitian symmetry, which possesses the same capacity for spontaneous symmetry transitions as the well-studied \(\mathcal{PT}\)-symmetry~\cite{Bender1998}. Thus, a pair of modes may be pulled together by coupling, colliding in an exceptional point where they transition from possessing \(\mathcal{RC}\)-symmetry only as a pair to each individually preserving it, with each lying directly on the symmetry axis \(A_{n}\). (d) An example of (c) in the FQNM trajectories of a Drude metal (parameters given below Eq. (\ref{eq:ExampleDrudeModel})) about the \(A_1\) symmetry axis. These trajectories are plotted as a function of length between \(L = 26\ c\Omega^{- 1}\) and \(L = 34\ c\Omega^{- 1}\).
    }
    \label{fig:RCExceptionalPointDemo}
\end{figure}

However, the effect of this coupling on mode trajectories is heavily constrained by symmetry, resulting in 3 distinct behaviours, represented schematically their respective panels of Figure \ref{fig:RCExceptionalPointDemo}.

In the first case we have an \emph{unperturbed crossing}. Modes cross the symmetry axis $A_n$ with a smooth path, their coupling begin zero on $A_n$. Since this requires the coupling to be exactly zero, this is the least likely to be observed in a given setup.  Second we have \emph{avoided crossing}. Interaction between the two modes prevents their collision entirely, and neither touches $A_n$.  Finally we have exceptional point crossing. Two modes attract as they approach $A_n$, rapidly colliding in an exceptional point, before rapidly splitting again into a pair lying along $A_n$. This process is then undone to allow the pair to leave $A_n$.

Out of these possibilities, of particular interest is case (c), as the rapid collision of two modes at an exceptional point is a non-perturbative phenomenon which is commonly observed (due to \(\mathcal{RC}\) symmetry) in the behaviour of FQNMs (see \ref{fig:RCExceptionalPointDemo}.d). This behaviour is closely analogous to the $\mathcal{PT}$-symmetry of~\cite{Bender1998}, with exceptional points occurring when the modes transition from being symmetric as a pair about \(A_n\), to being individually symmetric and lying on \(A_n\).  This symmetry transition also has direct physical consequences.  Modes away from a symmetry-axis \(A_{n}\), represent fields where the oscillation frequency does not have a fixed phase relationship with the modulation of the material parameters.  As a result, the exponential decay rate for these modes remains the same regardless of any relative offset between oscillations in a mode and modulation of the slab\footnote{Over shorter timescales, especially for frequencies which are very nearly commensurate, some initial amplification or decay may occur, before eventually the mode shifts out of phase with the driving field. For further discussion, see~\cite{Kiorpelidis2024}.}. By contrast, modes lying precisely on an axis \(A_{n}\) are phase-locked with the driving field. Furthermore, their phase relative to the driving field immediately determines whether the time modulation slows or expedites their decay. This phase sensitivity is a key feature of time-crystals, and has been discussed by various authors~\cite{Hooper2025, Galiffi2024, Hendry2025, Wang2025}.

We have thus demonstrated that the phase sensitivity is not a feature solely of bulk time-crystals, but that it also arises within finite time crystals. In addition, by viewing this through the language of symmetry transitions, we can see that this feature is in fact a robust feature of time-varying media.

\section{\label{sec:LargeCavities}Limiting Structure of Modes in Large Slabs}

The above has considered how time-modulation affects the motion of relatively few modes in the complex plane.   However, taking the limit of large slabs allows us to connect propagation in the bulk of a time--varying medium and the FQNMs of a finite slab, which are more relevant to typical experiments.  In particular, we demonstrate that the complex frequency associated with the maximum gain in a wavenumber bandgap is realised in finite slabs as a limit point in the set of FQNMs as slab length is increased to infinity.

\begin{figure}
    \centering
    \includegraphics[width=0.9\linewidth]{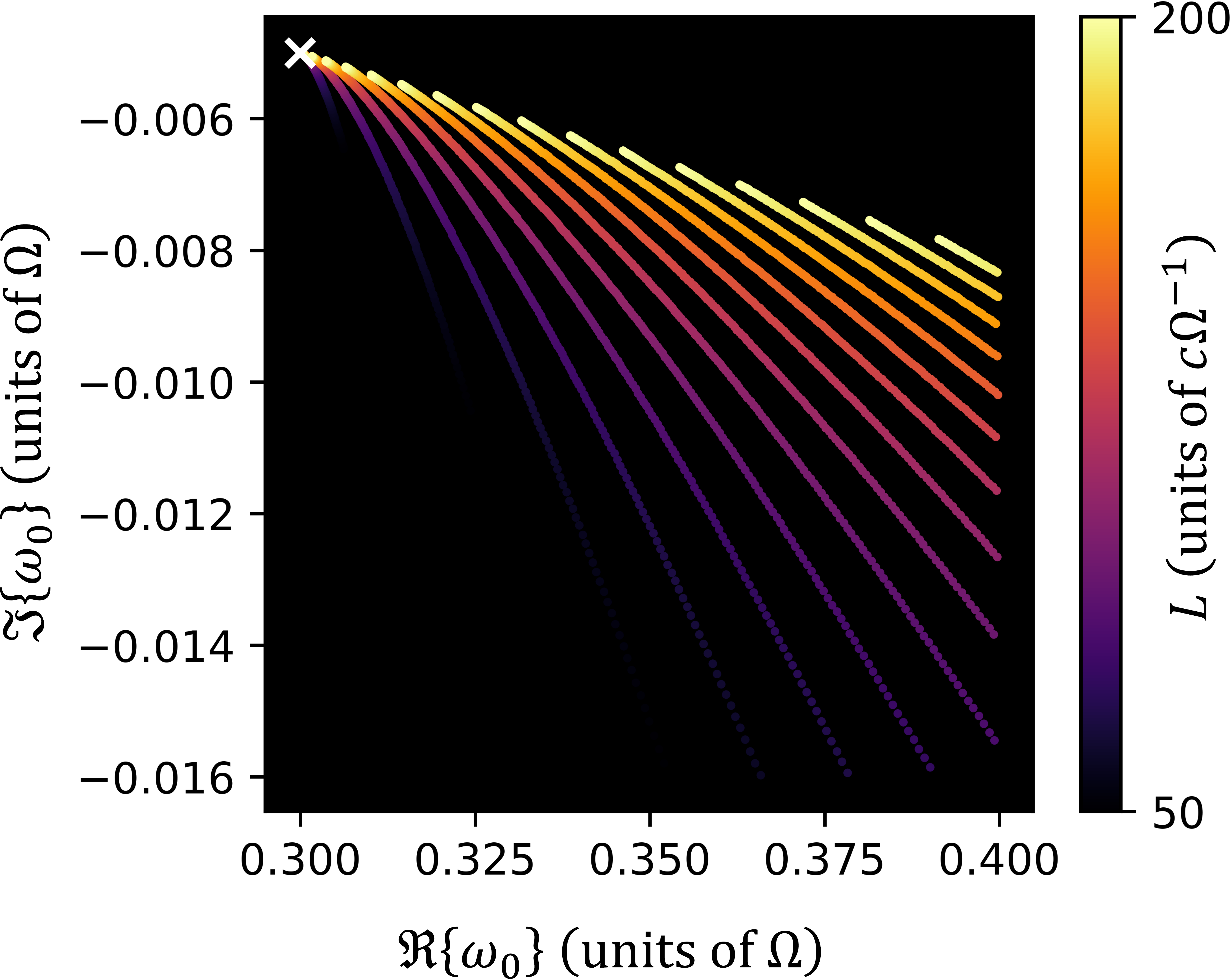}
    \caption{
        \textbf{The QNMs of a Drude slab approaching a limit point as slab length increases:}
        Plotted as points: sampled trajectories of QNMs for slab lengths increasing from \(L = 50\ c\Omega^{- 1}\) to \(L = 200\ c\Omega^{- 1}\). The white cross corresponds to a zero wavenumber point of the Drude model (discussed further in the main text), apparently attracting the QNMs of the system.
    }
    \label{fig:StaticStagnationDemo}
\end{figure}

Figure~\ref{fig:StaticStagnationDemo} shows a simple example for the case of a static medium, where many modes approach a single limiting complex frequency as \(L\to\infty\).  In static media, FQNM frequencies may be obtained directly from the equivalent QNM condition.  Replacing operators with scalars, Eq. (\ref{eq:FQNMCondition}) reduces to

\begin{equation}
    \left( 1 - \left( r(\omega) {\rm e}^{\mathrm{i}k(\omega)L}\right)^{2}\right) = 0,
    \label{eq:RoundTripQNMCondition}
\end{equation}

\noindent where \(k(\omega)\) and \(r(\omega)=\frac{1 - n(\omega)}{1 + n(\omega)}\) are the frequency dependent wavenumber and reflectivity of a static system with refractive index \(n(\omega)\).

The limiting frequency evident in Fig.~\ref{fig:StaticStagnationDemo} corresponds to the point \(\omega_{n=0}\) where the permittivity vanishes: the uniform mode of the slab remains at a fixed frequency, and other increasingly long wavelength modes tending towards this complex frequency with increasing $L$.  In the limit of an infinitely long slab, this ever more closely spaced set of QNMs will all satisfy the bulk dispersion relation, $k=\omega\sqrt{\epsilon(\omega)}/c$, parameterized by some complex value of $k$. 

To see this clustering of QNM frequencies directly, we take frequencies surrounding the zero index point, \(k^{2}\left( \omega_{n=0} \right) = 0\), writing the frequency as \(\omega = (1 + \delta)\omega_{n=0}\) and  \(k(\omega) \approx \mathcal{K}\sqrt{\delta}\), for \(\left|\delta\right|\ll1\). Note that, as \(L \rightarrow \infty\), \(k(\omega)L\) varies significantly faster than \(r(\omega)\), so the latter can be well approximated as unity. More specifically, our approximation is to take the \(\delta\to0\) limit, although allowing \(L\) to increase sufficiently that \(k(\omega)L\) does not decay to \(0\).

Under these conditions, the solutions to (\ref{eq:RoundTripQNMCondition}) can be approximated as,

\begin{equation}
    \omega \approx \omega_{n=0}\left( 1 + \left( \frac{ m\pi}{\mathcal{K}L} \right)^{2} \right)~\forall m\mathbb{\in Z,}
    \label{eq:QNMStagnationPoints}
\end{equation}

\noindent where the quadratic dependence on \(m\) is responsible for the clustering of modes around the zero index frequency identified in Fig. \ref{fig:StaticStagnationDemo}.  Taking the $L\to\infty$ limit and replacing $m$ with the continuous variable $k=m\pi/L$ this becomes,
\begin{equation}
    \omega\to\omega_{n=0}+\frac{c^2k^2}{\epsilon'(\omega_{\rm S})\omega_{n=0}^2}\label{eq:infinite-L-limit},
\end{equation}
where $\mathcal{K}=\sqrt{\epsilon'(\omega_{\rm S})}\,\omega_{n=0}^{3/2}/c$.  Eq. (\ref{eq:infinite-L-limit}) is just the bulk dispersion relation for a frequency close to the zero index point, parameterized by a real wave--vector $k$.

\subsection{\label{subsec:LargeTimeCrystalCavities}FQNM limits of a time--varying slab}

\subsubsection{\label{subsubsec:1DSubspace}The single mode case}

How does this behaviour of the QNM spectrum carry over to the case of a time--varying slab?  To answer this we first consider the analogue of a zero index point, a complex Floquet frequency, $\omega_{n=0}$ where a single eigenvalue of \({\widehat{K}}^{2}(\omega_{n=0})\) vanishes.  In the case of time--varying media this point represents the analogue of the uniform slab mode: a wave within the slab that has fixed zero wave--vector, and is composed of a combination of frequencies with the same relative amplitude after a single period of modulation.  As we shall show, for such time--varying systems there are also a set of FQNMs with different Floquet frequencies that, with increasing $L$, approach the point of zero index, where \({\widehat{K}}^{2}(\omega_{n=0})=0\).

To find the complex frequency $\omega_0$ of one of these modes we use Eqns. (\ref{eq:FQNMCondition}) and (\ref{eq:Q-kernel}), pre--multiplying $\widehat{Q}$ to write the FQNM condition as
\begin{equation}
    \left[1-\left(\widehat{r}{\rm e}^{{\rm i}\widehat{K}L}\right)^2\right]|\tilde{E}_{\rm int}\rangle=0
    \label{eq:qnm-operator-one}
\end{equation}
where $|\tilde{E}_{\rm int}\rangle=(\widehat{n}^{-1}-1)|\tilde{E}_{\rm out,+}\rangle$, and \(\widehat{r}=(1-\widehat{n})(1+\widehat{n})^{-1}\) as the time-varying analogue to the static reflection operator.

Choosing the root \(\widehat{K} \approx [{\widehat{K}}^{2}]^{1/2}\) such that eigenvalues of \({\rm e}^{\mathrm{i}\widehat{K}L}\) decay with increasing \(L\) eliminates all other modes from (\ref{eq:qnm-operator-one}) except those which---as in Eq. (\ref{eq:infinite-L-limit})---correspond to a real eigenvalue of $\widehat{K}$.  Here we assume that there is only one such eigenvalue (with corresponding eigenvector $|0\rangle$) for each Floquet frequency $\omega_0$, which thus turns the propagation operator into a projector, \({\rm e}^{\mathrm{i}\widehat{K}L}\approx{\rm e}^{{\rm i}k_0(\omega_0)L}|0\rangle\langle0|\).

Using the same limit as before for our approximation, where \(\delta\to0\) unless multiplied by the very large \(L\), the eigenvector \(|0\rangle\) becomes essentially constant, with \(\widehat{r}|0\rangle\approx|0\rangle\). Thus, being now equivalent to a $1\times1$ matrix problem, Eq. (\ref{eq:qnm-operator-one}) reduces to the scalar form

\begin{equation}
    \left( 1 - {\rm e}^{2\mathrm{i}k_{0}(\omega_0)L} \right)|\tilde{E}_{\mathrm{int}}\rangle \approx 0,
    \label{eq:DominatedRoundTripFQNMCondition}
\end{equation}

\noindent where the eigenvector is $|\tilde{E}_{\mathrm{int}}\rangle=|0\rangle$, and \(k_0\) is the relevant real eigenvalue of $\widehat{K}$.

Equation (\ref{eq:DominatedRoundTripFQNMCondition}) is equivalent to our earlier condition (\ref{eq:RoundTripQNMCondition}) for static media. Thus, provided only a single eigenvalue of the $\widehat{K}$ operator is real valued as a function of the Floquet frequency $\omega_0$, we will have the same cluster of FQNMs, approaching the zero index point with increasing slab length.

Since these limit points are qualitatively identical to those of the previous section, we term both ``static'' limit points.

\subsubsection{\label{subsubsec:2DSubspace}The two mode case}

The gain associated with the wavenumber bandgap of a time crystal arises from complex quasifrequencies \(\omega_0\) which are despite their complex nature are associated with a real wavenumber.  Mathematically, this corresponds to investigating when \(\widehat{K}^2\left(\omega_0\right)\) possesses real eigenvalues. Fortunately, the \(\mathcal{RC}\)-symmetry noted in Section~\ref{sec:FloquetQNMs} for \(\omega_0\) lying on a symmetry axis, is directly analogous~\cite{Hooper2025} to the \(\mathcal{PT}\)-symmetry investigated for non-Hermitian Hamiltonians in quantum mechanics~\cite{wang20132}, where the question of when an operator possesses real eigenvalues is well-studied in terms of symmetric and symmetry-broken phases.

In our case, whenever \(\omega_0\) lies on a symmetry axis \(A_n\), \(\widehat{K}^2\left(\omega_0\right)\) may fall into either of these phases---a symmetric phase with real eigenvalues, or a symmetry-broken phase where eigenvalues are found in complex-conjugate pairs---separated by an exceptional point where the operator cannot be diagonalised. As such, the wavenumber bandgap in a time crystal corresponds precisely to this symmetric phase, with the maximum gain possible in determined by the maximum \(\Im\left\{\omega_0\right\}\) in the symmetric phase, and thus by the position of the exceptional point marking the end of such a phase.

Such exceptional points, naturally requiring the coupling between two frequencies, are inherently time-varying in nature. And, as we will demonstrate, form limit points precisely analogous to those of the previous section.

To demonstrate this, we return to our methods of the previous sections: consider frequencies \(\omega_0=\left(1+\delta\right)\omega_{\rm EP}\) around an exceptional point \(\omega_{\rm EP}\) where a pair of eigenvalues of \(\widehat{K}^2\) transition from their symmetric (real) phase, to their symmetry-broken (complex conjugate pair) phase. Characteristic of an exceptional point is that this collision does not occur linearly in \(\delta\), but with a square root dependence -- a dependence inherited by the eigenvalues of \(\widehat{K}=[\widehat{K}^2]^{1/2}\). This square root collision can be written explicitly by considering writing the eigenvalues \(k_\pm(\omega_0)\) in terms of their average \(\overline{k}\) and half-splitting \(\Delta k\), such that \(k_\pm(\omega_0)=\overline{k}(\omega_0)\pm\Delta k(\omega_0)\), which, around the exceptional point (to first order in \(\sqrt\delta\)), reduces to

\begin{equation}
    k_\pm(\omega_0)\approx\overline{k}(\omega_{\rm EP})\pm\Delta \mathcal{K}\sqrt{\delta}.
    \label{eq:ExceptionalPointEigenvalueCollision}
\end{equation}

Of course, any eigenvectors associated with the symmetric phase cannot decay in magnitude under evolution by \({\rm e}^{{\rm i}\widehat{K}L}\). Thus, following our approach from the previous section, assume that all other modes of \(\widehat{K}^2\) contain at least a small amount of loss (although this assumption is actually unnecessary) such that, for large slab lengths \(L\), \({\rm e}^{{\rm i}\widehat{K}L}\) projects to \(0\) all but \(2\) modes.

It would be convenient to follow our prior analyses by finding a diagonalisation for \(\widehat{K}^2\) which holds to constant order in \(\sqrt{\delta}\), then representing \({\rm e}^{{\rm i}\widehat{K}L}\) using this basis. However, \(\delta=0\) corresponds, by definition, to an exceptional point of \(\widehat{K}^2\) where diagonalisation is impossible. Instead, we apply the Schur decomposition, to at least write \(\widehat{K}^2\) in upper diagonal form.

We thus consider only a single eigenvector \(\widehat{K}^2|+\rangle=k_+^2|+\rangle\), letting \(\langle+|=(|+\rangle)^\dagger\), normalised as for a standard orthonormal basis, before introducing the second vector \(|-\rangle\) to span the remaining space. However, in contrast to diagonalisation, \(|-\rangle\) is not an eigenvector of \(\widehat{K}^2\), but instead defined by its orthonormality to \(|+\rangle\), with \(\langle-|=(|-\rangle)^\dagger\), \(\langle-|-\rangle=1\) and \(\langle+|-\rangle=0\).

In this basis, \(\widehat{K}^2\) is then written as
\begin{equation}
    \widehat{K}^2=
    k_+^2|+\rangle\langle+|+\kappa^2|+\rangle\langle-|+k_-^2|-\rangle\langle-|,
    \label{eq:SchurK2}
\end{equation}
with \(\kappa^2=\langle+|\widehat{K}^2|-\rangle\) as a dimensionful quantity encoding the extent to which \(|-\rangle\) fails to be an eigenvalue of \(\widehat{K}^2\).

Sylvester's formula for functions of $2\times 2$ matrices then allows us to explicitly find \({\rm e}^{{\rm i}\widehat{K}L}\) as we approach the exceptional point:

\begin{equation}
    \begin{split}
        {\rm e}^{{\rm i}\widehat{K}L} = {\rm e}^{{\rm i}\overline{k}L} & \left( {\rm e}^{{\rm i}\Delta kL}\frac{\widehat{K}^2-k_-^2}{k_+^2-k_-^2}\right.\\
        &-\left.{\rm e}^{-{\rm i}\Delta kL}\frac{\widehat{K}^2-k_+^2}{k_+^2-k_-^2}\right).
    \end{split}
    \label{eq:SylvesterExponentialFormula}
\end{equation}

However, about the exceptional point, \(k_+^2-k_-^2=4\overline{k}\Delta k\) tends to zero as \(\sqrt\delta\). To assess the resulting divergence, the projectors in Sylvester's formula may be written in terms of the \(|\pm\rangle\) basis:

\begin{equation}
    \frac{\widehat{K}^2-k_\pm^2}{k_+^2-k_-^2}= \mp|\mp\rangle\langle\mp|+\frac{\kappa^2}{4\overline{k}\Delta k}|+\rangle\langle-|.
    \label{eq:DivergentProjectors}
\end{equation}

For fixed \(L\), this divergence is canceled by the fact that \({\rm e}^{{\rm i}\Delta kL}\to1\). However, for arbitrarily large \(L\), and small \(\delta\), this cancellation does not occur, and as \(\delta\to0\) the exponential is dominated by

\begin{equation}
    {\rm e}^{{\rm i}\widehat{K}L}\approx{\rm e}^{{\rm i}\overline{k}L}\frac{{\rm i}\kappa^2}{2\overline{k}\Delta k}\sin\left(\Delta k\,L\right)|+\rangle\langle-|.
\end{equation}

Thus, in this limit, Equation (\ref{eq:qnm-operator-one}) tends towards the form,

\begin{equation}
    -\left(\widehat{r}{\rm e}^{{\rm i}\overline{k}L}\frac{{\rm i}\kappa^2}{2\overline{k}\Delta k}\sin\left(\Delta k\,L\right)|+\rangle\langle-|\right)^2|\tilde{E}_{\rm int}\rangle\approx0.
    \label{eq:DefectiveFQNMCondition}
\end{equation}

By letting \(|\tilde{E}_{\rm int}\rangle=\widehat{r}|+\rangle\), and dividing out constants in \(L\), this then reduces to
\begin{equation}
    \left( 1 - {\rm e}^{2\mathrm{i}\Delta k(\omega_0)L} \right)^2|\tilde{E}_{\mathrm{int}}\rangle \approx 0,
    \label{eq:DefectiveRoundTripFQNMCondition}
\end{equation}
which, in precise analogy with both Equations (\ref{eq:RoundTripQNMCondition}) and (\ref{eq:DominatedRoundTripFQNMCondition}), this has double roots at
\begin{equation}
    \omega\approx\omega_{\rm EP}\left(1+\left(\frac{m\pi}{\Delta\mathcal{K}L}\right)^2\right)~\forall m\in\mathbb{Z}.
    \label{eq:FloquetLimitPointApproach}
\end{equation}
The effect of higher order corrections in \(\delta\) is then simply to split these pairs of roots by a small amount.

Given that the origin of these limit points is qualitatively distinct from the ``static'' limit points described above -- arising only due to coupling between different frequencies in a time-varying medium -- we henceforth refer to these modes as ``exceptional'' limit points.

\begin{figure}
    \centering
    \includegraphics[width=0.9\linewidth]{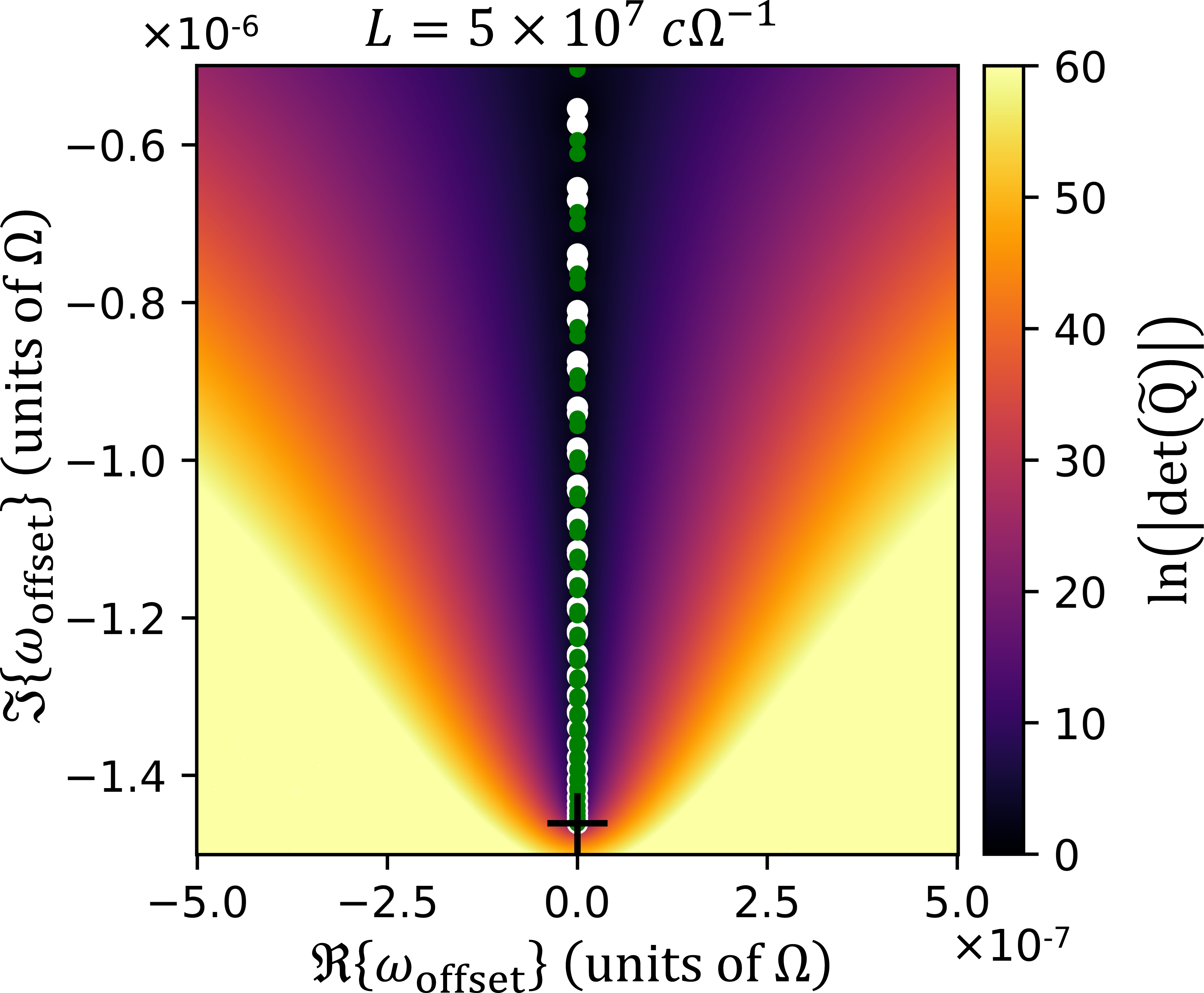}
    \caption{
        \textbf{Exceptional limit point in a large slab:}
        For illustrative purposes, in this example, rather than a Drude model, we consider the \(4 \times 4\) truncation of \(\widehat{\chi} = 1 + 0.2{\widehat{\Delta}}_{1}(0)\). We plot the FQNM indicator function as a colourmap, with FQNM quasifrequencies highlighted by white dots, and their approximations plotted in smaller green dots. All frequencies are plotted relative to the \(\mathcal{RC}\)-symmetry axis \(A_{1}\) with \(\omega_\mathrm{offset}=\omega_0-\frac{\Omega}{2}\).
        For a very thick slab (\(L = 5 \times 10^{7}\ c\Omega^{- 1}\)), many modes have clustered around this exceptional limit point, associated with the weak third wavenumber bandgap in the dispersion relation of \(\widehat{K}^2(\omega_0)\).
        Indeed, our approximation remains valid despite the wavenumber bandgap under consideration actually lying entirely within (in terms of its complex frequency content), the much stronger first wavenumber bandgap.
    }
    \label{fig:TimeCrystalStagnation}
\end{figure}

Note that this divergence associated with the defective point \(\delta\to0\) actually makes our earlier assumption that \({\rm e}^{{\rm i}\widehat{K}L}\) becomes a \(2\times2\) projector unnecessary. The defective point naturally dominates as long as the root \(\widehat{K}=[\widehat{K}^2]^{1/2}\) is chosen we stated: so that \({\rm e}^{{\rm i}\widehat{K}L}\) is never exponentially growing in \(L\). In Fig.~\ref{fig:TimeCrystalStagnation}, we demonstrate the accuracy of our analysis by approximating the FQNMs of a large slab by the roots of \(\det\left(\sin\left(\widehat{K}\left(\omega_0\right)L\right)\right)=0\). This example confirms our analysis even in the case that \({\rm e}^{{\rm i}\widehat{K}L}\) does not project down onto a \(2\times2\) subspace.

This analysis demonstrates the clustering of modes around the exceptional point frequencies \(\omega_{\rm EP}\) associated with the maximum and minimum loss in the wavenumber bandgaps of a system. The simple approximation presented, also bears direct relevance to experiments, directly relating the thickness of a slab to how well it realises the gain predicted by the wavenumber bandgap of its bulk.

\section{\label{sec:Results}Modes of a Time-Varying Drude Slab}

In this section, we demonstrate the accuracy of our analyses by returning to our Drude model example in the case of a very thick slab. We note that our results identify the origin of the diverging transmission coefficients we noted previously in~\cite{Hooper2025}. These poles arise whenever an FQNM crosses the real frequency axis. Indeed, such crosses are guaranteed by the wavenumber bandgap in this material, which places an exceptional limit point above the real axis, directly at the point of maximum gain of the bulk system.

In more generality, to assess the validity of our analyses throughout this paper, we note the following predictions made over the preceding sections, as applied to a slab of increasing length:

\begin{enumerate}
    \item\label{item:SmoothTrajectories} The trajectories followed by any given FQNM will be largely smooth.
    \item\label{item:Collisions} An exception to statement \ref{item:SmoothTrajectories} occurs whenever two modes collide along a symmetry axis \(A_n\), resulting in a pair of exceptional points when the modes enter and leave \(A_n\).
    \item\label{item:Stagnation} Modes will tend towards both static and exceptional limit points, as defined in the previous section, associated respectively with refractive index zeros and the (complex frequency) limits of wavenumber bandgaps.
\end{enumerate}

\begin{figure}
    \centering
    \includegraphics[width=0.9\linewidth]{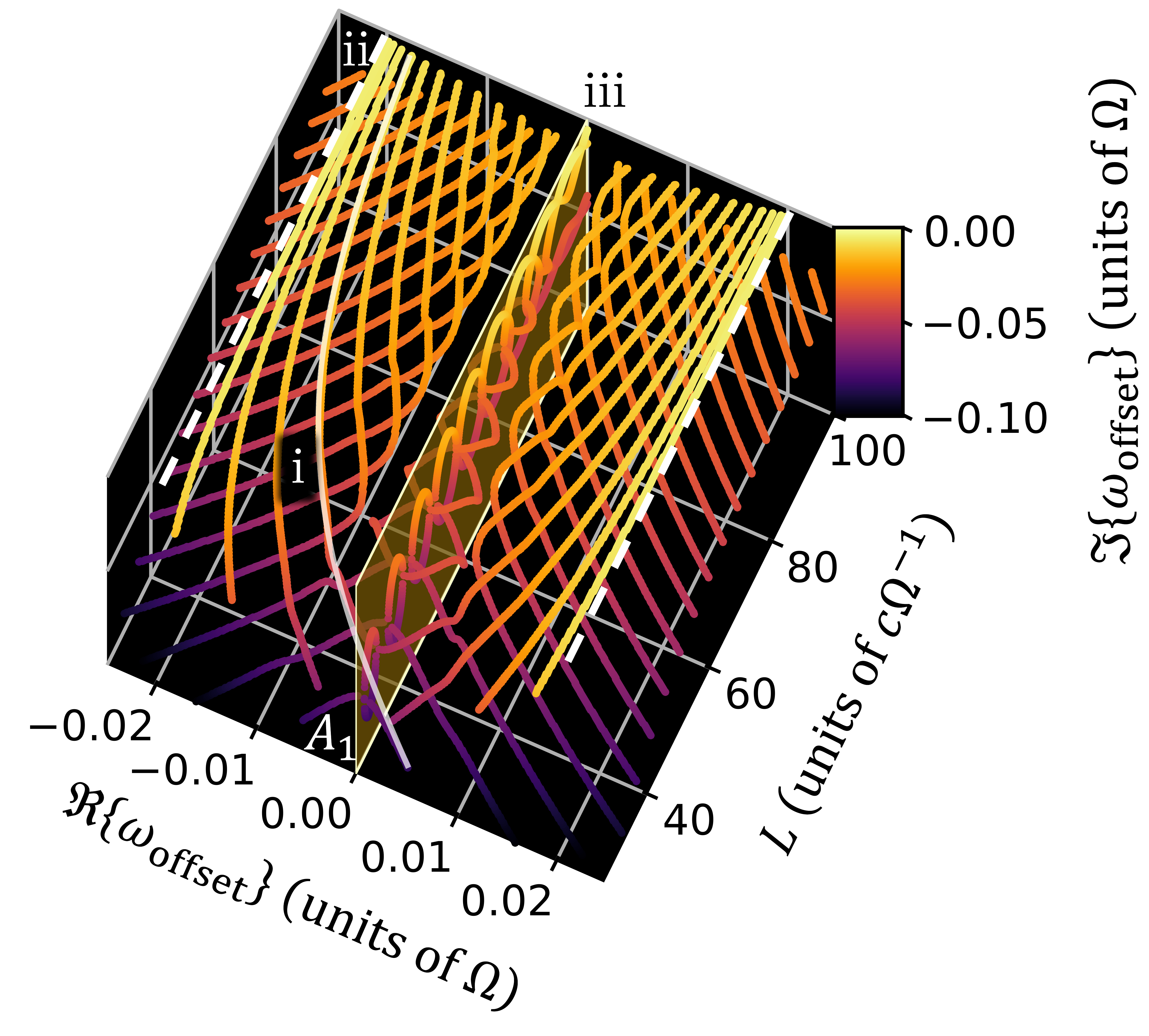}
    \caption{
        \textbf{FQNMs in a Drude slab for a large range of increasing lengths:}
        Sampled FQNM trajectories are plotted as points coloured by the imaginary part of their quasifrequency. For readability, only modes for a \(2\times2\) truncation of \(\widehat{\chi}\) are plotted. The \(\mathcal{RC}\)-symmetry axis \(A_{1}\) is shown as a translucent yellow plane, with a light yellow outline. (i) An otherwise smooth FQNM trajectory (qualitatively presented with a transparent white line) in a static system is broken in time-varying media by \(\mathcal{RC}\)-symmetry transitions about the \(A_{1}\) axis. (ii) Most FQNM trajectories have limiting behaviour similar to static systems, approaching refractive index zeros (white dashed lines) as slab length increases. (iii) However, for sufficiently thick slabs, pairs of modes become permanently confined to \(A_1\) and isolated from the static limit points (ii). These modes are associated closely with the bulk wavenumber bandgap, and responsible for the diverging transmission coefficients of the slabs presented in~\cite{Hooper2025}. Noting their isolation, one might predict that they tend towards an exceptional limit point, which is validated when considering much larger slabs, as in Figure~\ref{fig:RCStagnationPoint}.
    }
    \label{fig:DrudeFQNMs}
\end{figure}

In particular, we focus on frequencies around the \(\mathcal{RC}\)-symmetry axis \(A_1\). We thus consider the frequency variable \(\omega_{\rm offset} = \omega_{0} - \frac{\Omega}{2}\), and plot these FQNMs in Figure \ref{fig:DrudeFQNMs}.

Statements \ref{item:SmoothTrajectories} and \ref{item:Collisions} are confirmed by trajectories such as that feature (i) of Figure~\ref{fig:DrudeFQNMs}, which progresses smoothly aside from at \(A_1\), where two modes collide and then split in a pair of exceptional points, similar to those seen in Figure~\ref{fig:RCExceptionalPointDemo}.d. However, aside from collisions on \(A_1\), all trajectories remain continuous.

Meanwhile, statement \ref{item:Stagnation} is backed up most obviously by Figure~\ref{fig:DrudeFQNMs}.ii, as most of the modes pictured approach a static limit point. However, this is not true of every mode (see Figure~\ref{fig:DrudeFQNMs}.iii). In addition, two pairs of modes collide in an exceptional point, and remain stuck to the symmetry axis \(A_1\). This pairing up is precisely that expected of modes tending towards an exceptional limit point.

\begin{figure}
    \centering
    \includegraphics[width=0.9\linewidth]{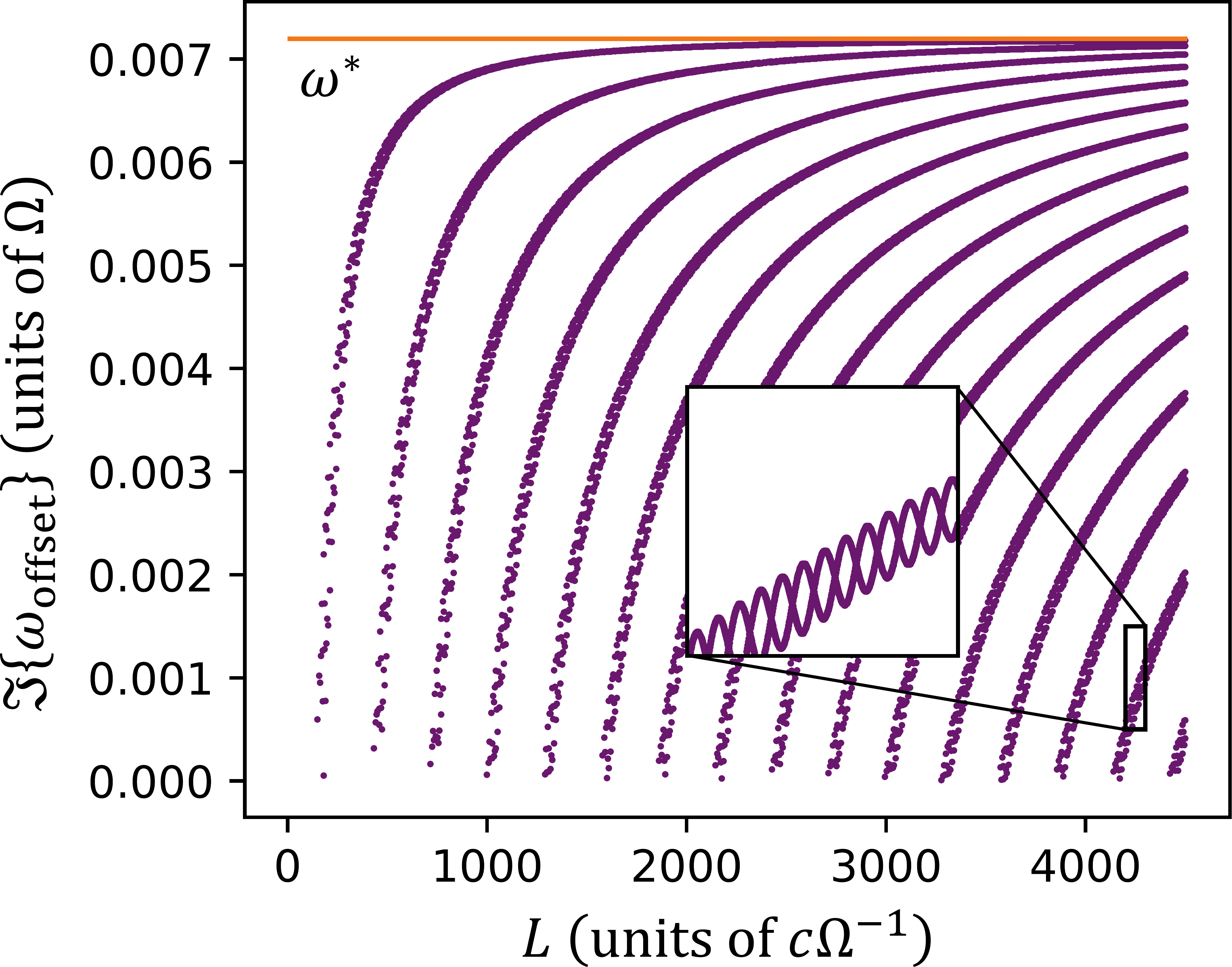}
    \caption{
        \textbf{An exceptional limit point due to \(\mathcal{RC}\)-symmetry in a Drude metal slab:}
        The sampled FQNM trajectories of a Drude metal slab converge to an exceptional limit point at \(\omega^{*}\). as its length increases from \(L=10\ c\Omega ^{-1}\) to \(L=4500\ c\Omega ^{-1}\).
        Eventually additional modes join the pairs shown in Figure \ref{fig:DrudeFQNMs}.iii in becoming confined to the \(A_1\) axis (pictured here), and over this large range of slab lengths, these converge slowly towards the limit point \(\omega^{*}\), associated with the maximum gain in the bulk wavenumber bandgap.
        In the inset we demonstrate that the apparent noise in the first half of the figure is really just a sampling artefact as pairs of orthogonal modes oscillate very rapidly.
    }
    \label{fig:RCStagnationPoint}
\end{figure}

Indeed, an exceptional point of \(\widehat{K}^2\) can be found, associated with \(\mathcal{RC}\)-symmetry breaking, at \({\omega^*\sim 0.0072\Omega \mathrm{i} + \frac{\Omega}{2}}\), predicting this as a limit point at the maximum gain available in a bulk wavenumber bandgap. In Figure~\ref{fig:RCStagnationPoint}, we demonstrate that this analysis is indeed correct: as the length of the Drude slab increases, ever more modes become confined to \(A_1\), oscillating in pairs whilst tending towards \(\omega^*\).

In Figure~\ref{fig:RCStagnationPoint}, this prediction is confirmed: as the length of the Drude slab increases, ever more modes become confined to \(A_1\), tending towards an exceptional limit point associated with \(\mathcal{RC}\)-symmetry breaking in \(\widehat{K}^2\). As discussed in the previous section, this exceptional point corresponds precisely to the maximum gain possible for a real wavenumber excitation in a bulk medium.

Furthermore, since the modes associated with the onset of gain are trapped to the axis \(A_1\) by \(\mathcal{RC}\)-symmetry, it follows that no sequence of smooth perturbations through real physical systems (which must preserve \(\mathcal{RC}\)-symmetry) could have predicted the onset of this gain.

\section{\label{sec:Conclusion}Summary and conclusions}
Over the course of this paper, we have introduced the notion of Floquet quasinormal modes (FQNMs), solutions to the wave equation in a periodically modulated cavity with outgoing boundary conditions with a definite Floquet quasifrequency. These FQNMs allow for a transient analysis of photonic time-crystal slabs, extending the driven response we investigated previously in~\cite{Hooper2025}. We note that collisions between FQNMs are not only guaranteed by symmetry, but that these result in exceptional points where pairs of FQNMs undergo a symmetry transition, becoming temporarily phase-locked to the modulation of the slab.

By directing our focus towards understanding how the FQNMs of a system vary under continuous changes to said system, we uncover a number of general properties of their trajectories in the complex plane. Indeed, we show that whilst these trajectories are typically smooth, FQNMs regularly collide along particular symmetry axes, forming exceptional points which interrupt this smoothness. Furthermore, by considering the limit points of these trajectories, we show that such collisions are essentially inevitable. In particular, we demonstrate that such non-perturbative collisions are actually required to understand the modes which realise gain in media with a wavenumber bandgap.

This non-perturbative approach contrasts other recent work in this area~\cite{Valero2025, Vial2025} by applying an operator-based approach~\cite{Horsley2023Operators} to focus primarily on non-perturbative features of the FQNM distribution. This focus allows us to identify the pitfalls to be encountered by future practical applications of perturbative approaches. In particular, due to the non-perturbative nature of collisions, we anticipate that any future perturbation theories aiming to provide convergent results must carefully select an initial model from which to perturb. In particular, our results suggest that direct perturbative analyses will typically not converge unless beginning from time-varying, rather than static, cavities.

\begin{acknowledgments}
CMH acknowledges financial support from the Engineering and Physical Sciences Research Council (EPSRC) of the UK via the Exeter University Physics DTP. IRH and SARH acknowledge financial support from the EPSRC via the META4D Programme Grant (EP/Y015673/1). SARH thanks the Royal Society and TATA for financial support (RPG-2016-186).
\end{acknowledgments}

\bibliography{apssamp}

\providecommand{\noopsort}[1]{}\providecommand{\singleletter}[1]{#1}%
\begin{thebibliography}{41}%
\makeatletter
\providecommand \@ifxundefined [1]{%
 \@ifx{#1\undefined}
}%
\providecommand \@ifnum [1]{%
 \ifnum #1\expandafter \@firstoftwo
 \else \expandafter \@secondoftwo
 \fi
}%
\providecommand \@ifx [1]{%
 \ifx #1\expandafter \@firstoftwo
 \else \expandafter \@secondoftwo
 \fi
}%
\providecommand \natexlab [1]{#1}%
\providecommand \enquote  [1]{``#1''}%
\providecommand \bibnamefont  [1]{#1}%
\providecommand \bibfnamefont [1]{#1}%
\providecommand \citenamefont [1]{#1}%
\providecommand \href@noop [0]{\@secondoftwo}%
\providecommand \href [0]{\begingroup \@sanitize@url \@href}%
\providecommand \@href[1]{\@@startlink{#1}\@@href}%
\providecommand \@@href[1]{\endgroup#1\@@endlink}%
\providecommand \@sanitize@url [0]{\catcode `\\12\catcode `\$12\catcode `\&12\catcode `\#12\catcode `\^12\catcode `\_12\catcode `\%12\relax}%
\providecommand \@@startlink[1]{}%
\providecommand \@@endlink[0]{}%
\providecommand \url  [0]{\begingroup\@sanitize@url \@url }%
\providecommand \@url [1]{\endgroup\@href {#1}{\urlprefix }}%
\providecommand \urlprefix  [0]{URL }%
\providecommand \Eprint [0]{\href }%
\providecommand \doibase [0]{https://doi.org/}%
\providecommand \selectlanguage [0]{\@gobble}%
\providecommand \bibinfo  [0]{\@secondoftwo}%
\providecommand \bibfield  [0]{\@secondoftwo}%
\providecommand \translation [1]{[#1]}%
\providecommand \BibitemOpen [0]{}%
\providecommand \bibitemStop [0]{}%
\providecommand \bibitemNoStop [0]{.\EOS\space}%
\providecommand \EOS [0]{\spacefactor3000\relax}%
\providecommand \BibitemShut  [1]{\csname bibitem#1\endcsname}%
\let\auto@bib@innerbib\@empty
\bibitem [{\citenamefont {Noether}(1918)}]{Noether1918}%
  \BibitemOpen
  \bibfield  {author} {\bibinfo {author} {\bibfnamefont {E.}~\bibnamefont {Noether}},\ }\bibfield  {title} {\bibinfo {title} {Invariante variationsprobleme},\ }\href {http://eudml.org/doc/59024} {\bibfield  {journal} {\bibinfo  {journal} {Nachrichten von der Gesellschaft der Wissenschaften zu Göttingen, Mathematisch-Physikalische Klasse}\ }\textbf {\bibinfo {volume} {1918}},\ \bibinfo {pages} {235} (\bibinfo {year} {1918})}\BibitemShut {NoStop}%
\bibitem [{\citenamefont {Galiffi}\ \emph {et~al.}(2022)\citenamefont {Galiffi}, \citenamefont {Tirole}, \citenamefont {Yin}, \citenamefont {Li}, \citenamefont {Vezzoli}, \citenamefont {Huidobro}, \citenamefont {Silveirinha}, \citenamefont {Sapienza}, \citenamefont {Al{\`u}},\ and\ \citenamefont {Pendry}}]{Galiffi2022}%
  \BibitemOpen
  \bibfield  {author} {\bibinfo {author} {\bibfnamefont {E.}~\bibnamefont {Galiffi}}, \bibinfo {author} {\bibfnamefont {R.}~\bibnamefont {Tirole}}, \bibinfo {author} {\bibfnamefont {S.}~\bibnamefont {Yin}}, \bibinfo {author} {\bibfnamefont {H.}~\bibnamefont {Li}}, \bibinfo {author} {\bibfnamefont {S.}~\bibnamefont {Vezzoli}}, \bibinfo {author} {\bibfnamefont {P.~A.}\ \bibnamefont {Huidobro}}, \bibinfo {author} {\bibfnamefont {M.~G.}\ \bibnamefont {Silveirinha}}, \bibinfo {author} {\bibfnamefont {R.}~\bibnamefont {Sapienza}}, \bibinfo {author} {\bibfnamefont {A.}~\bibnamefont {Al{\`u}}},\ and\ \bibinfo {author} {\bibfnamefont {J.~B.}\ \bibnamefont {Pendry}},\ }\bibfield  {title} {\bibinfo {title} {{Photonics of time-varying media}},\ }\href {https://doi.org/10.1117/1.AP.4.1.014002} {\bibfield  {journal} {\bibinfo  {journal} {Advanced Photonics}\ }\textbf {\bibinfo {volume} {4}},\ \bibinfo {pages} {014002} (\bibinfo {year} {2022})}\BibitemShut {NoStop}%
\bibitem [{\citenamefont {Caloz}\ and\ \citenamefont {Deck-Léger}(2020)}]{Caloz2020}%
  \BibitemOpen
  \bibfield  {author} {\bibinfo {author} {\bibfnamefont {C.}~\bibnamefont {Caloz}}\ and\ \bibinfo {author} {\bibfnamefont {Z.-L.}\ \bibnamefont {Deck-Léger}},\ }\bibfield  {title} {\bibinfo {title} {Spacetime metamaterials—part ii: Theory and applications},\ }\href {https://doi.org/10.1109/TAP.2019.2944216} {\bibfield  {journal} {\bibinfo  {journal} {IEEE Transactions on Antennas and Propagation}\ }\textbf {\bibinfo {volume} {68}},\ \bibinfo {pages} {1583} (\bibinfo {year} {2020})}\BibitemShut {NoStop}%
\bibitem [{\citenamefont {Cartella}\ \emph {et~al.}(2018)\citenamefont {Cartella}, \citenamefont {Nova}, \citenamefont {Fechner}, \citenamefont {Merlin},\ and\ \citenamefont {Cavalleri}}]{Cartella2018}%
  \BibitemOpen
  \bibfield  {author} {\bibinfo {author} {\bibfnamefont {A.}~\bibnamefont {Cartella}}, \bibinfo {author} {\bibfnamefont {T.~F.}\ \bibnamefont {Nova}}, \bibinfo {author} {\bibfnamefont {M.}~\bibnamefont {Fechner}}, \bibinfo {author} {\bibfnamefont {R.}~\bibnamefont {Merlin}},\ and\ \bibinfo {author} {\bibfnamefont {A.}~\bibnamefont {Cavalleri}},\ }\bibfield  {title} {\bibinfo {title} {Parametric amplification of optical phonons},\ }\href {https://doi.org/10.1073/pnas.1809725115} {\bibfield  {journal} {\bibinfo  {journal} {Proceedings of the National Academy of Sciences}\ }\textbf {\bibinfo {volume} {115}},\ \bibinfo {pages} {12148} (\bibinfo {year} {2018})},\ \Eprint {https://arxiv.org/abs/https://www.pnas.org/doi/pdf/10.1073/pnas.1809725115} {https://www.pnas.org/doi/pdf/10.1073/pnas.1809725115} \BibitemShut {NoStop}%
\bibitem [{\citenamefont {Castaldi}\ \emph {et~al.}(2023)\citenamefont {Castaldi}, \citenamefont {Rizza}, \citenamefont {Engheta},\ and\ \citenamefont {Galdi}}]{Castaldi2023}%
  \BibitemOpen
  \bibfield  {author} {\bibinfo {author} {\bibfnamefont {G.}~\bibnamefont {Castaldi}}, \bibinfo {author} {\bibfnamefont {C.}~\bibnamefont {Rizza}}, \bibinfo {author} {\bibfnamefont {N.}~\bibnamefont {Engheta}},\ and\ \bibinfo {author} {\bibfnamefont {V.}~\bibnamefont {Galdi}},\ }\bibfield  {title} {\bibinfo {title} {Multiple actions of time-resolved short-pulsed metamaterials},\ }\href {https://doi.org/10.1063/5.0132554} {\bibfield  {journal} {\bibinfo  {journal} {Applied Physics Letters}\ }\textbf {\bibinfo {volume} {122}},\ \bibinfo {pages} {021701} (\bibinfo {year} {2023})},\ \Eprint {https://arxiv.org/abs/https://pubs.aip.org/aip/apl/article-pdf/doi/10.1063/5.0132554/16741904/021701\_1\_online.pdf} {https://pubs.aip.org/aip/apl/article-pdf/doi/10.1063/5.0132554/16741904/021701\_1\_online.pdf} \BibitemShut {NoStop}%
\bibitem [{\citenamefont {Zhu}\ \emph {et~al.}(2023)\citenamefont {Zhu}, \citenamefont {Wu}, \citenamefont {Zhuo}, \citenamefont {Liu},\ and\ \citenamefont {Li}}]{Zhu2023}%
  \BibitemOpen
  \bibfield  {author} {\bibinfo {author} {\bibfnamefont {X.}~\bibnamefont {Zhu}}, \bibinfo {author} {\bibfnamefont {H.-W.}\ \bibnamefont {Wu}}, \bibinfo {author} {\bibfnamefont {Y.}~\bibnamefont {Zhuo}}, \bibinfo {author} {\bibfnamefont {Z.}~\bibnamefont {Liu}},\ and\ \bibinfo {author} {\bibfnamefont {J.}~\bibnamefont {Li}},\ }\bibfield  {title} {\bibinfo {title} {Effective medium for time-varying frequency-dispersive acoustic metamaterials},\ }\href {https://doi.org/10.1103/PhysRevB.108.104303} {\bibfield  {journal} {\bibinfo  {journal} {Phys. Rev. B}\ }\textbf {\bibinfo {volume} {108}},\ \bibinfo {pages} {104303} (\bibinfo {year} {2023})}\BibitemShut {NoStop}%
\bibitem [{\citenamefont {{a}o C.~Serra}\ \emph {et~al.}(2024)\citenamefont {{a}o C.~Serra}, \citenamefont {Galiffi}, \citenamefont {Huidobro}, \citenamefont {Pendry},\ and\ \citenamefont {Silveirinha}}]{Serra2024}%
  \BibitemOpen
  \bibfield  {author} {\bibinfo {author} {\bibfnamefont {J.}~\bibnamefont {{a}o C.~Serra}}, \bibinfo {author} {\bibfnamefont {E.}~\bibnamefont {Galiffi}}, \bibinfo {author} {\bibfnamefont {P.~A.}\ \bibnamefont {Huidobro}}, \bibinfo {author} {\bibfnamefont {J.~B.}\ \bibnamefont {Pendry}},\ and\ \bibinfo {author} {\bibfnamefont {M.~G.}\ \bibnamefont {Silveirinha}},\ }\bibfield  {title} {\bibinfo {title} {Particle-hole instabilities in photonic time-varying systems},\ }\href {https://doi.org/10.1364/OME.521571} {\bibfield  {journal} {\bibinfo  {journal} {Opt. Mater. Express}\ }\textbf {\bibinfo {volume} {14}},\ \bibinfo {pages} {1459} (\bibinfo {year} {2024})}\BibitemShut {NoStop}%
\bibitem [{\citenamefont {Pendry}(2024)}]{Pendry2024}%
  \BibitemOpen
  \bibfield  {author} {\bibinfo {author} {\bibfnamefont {J.~B.}\ \bibnamefont {Pendry}},\ }\bibfield  {title} {\bibinfo {title} {Air conditioning for photons [invited]},\ }\href {https://doi.org/10.1364/OME.511182} {\bibfield  {journal} {\bibinfo  {journal} {Opt. Mater. Express}\ }\textbf {\bibinfo {volume} {14}},\ \bibinfo {pages} {407} (\bibinfo {year} {2024})}\BibitemShut {NoStop}%
\bibitem [{\citenamefont {Dong}\ \emph {et~al.}(2025)\citenamefont {Dong}, \citenamefont {Chen},\ and\ \citenamefont {Yuan}}]{Dong2025}%
  \BibitemOpen
  \bibfield  {author} {\bibinfo {author} {\bibfnamefont {Z.}~\bibnamefont {Dong}}, \bibinfo {author} {\bibfnamefont {X.}~\bibnamefont {Chen}},\ and\ \bibinfo {author} {\bibfnamefont {L.}~\bibnamefont {Yuan}},\ }\bibfield  {title} {\bibinfo {title} {Extremely narrow band in moir\'e photonic time crystal},\ }\href {https://doi.org/10.1103/4lqd-z567} {\bibfield  {journal} {\bibinfo  {journal} {Phys. Rev. Lett.}\ }\textbf {\bibinfo {volume} {135}},\ \bibinfo {pages} {033803} (\bibinfo {year} {2025})}\BibitemShut {NoStop}%
\bibitem [{\citenamefont {Lustig}\ \emph {et~al.}(2018)\citenamefont {Lustig}, \citenamefont {Sharabi},\ and\ \citenamefont {Segev}}]{Lustig2018}%
  \BibitemOpen
  \bibfield  {author} {\bibinfo {author} {\bibfnamefont {E.}~\bibnamefont {Lustig}}, \bibinfo {author} {\bibfnamefont {Y.}~\bibnamefont {Sharabi}},\ and\ \bibinfo {author} {\bibfnamefont {M.}~\bibnamefont {Segev}},\ }\bibfield  {title} {\bibinfo {title} {Topological aspects of photonic time crystals},\ }\href {https://doi.org/10.1364/OPTICA.5.001390} {\bibfield  {journal} {\bibinfo  {journal} {Optica}\ }\textbf {\bibinfo {volume} {5}},\ \bibinfo {pages} {1390} (\bibinfo {year} {2018})}\BibitemShut {NoStop}%
\bibitem [{\citenamefont {Harwood}\ \emph {et~al.}(2025)\citenamefont {Harwood}, \citenamefont {Vezzoli}, \citenamefont {Raziman}, \citenamefont {Hooper}, \citenamefont {Tirole}, \citenamefont {Wu}, \citenamefont {Maier}, \citenamefont {Pendry}, \citenamefont {Horsley},\ and\ \citenamefont {Sapienza}}]{Harwood2025}%
  \BibitemOpen
  \bibfield  {author} {\bibinfo {author} {\bibfnamefont {A.~C.}\ \bibnamefont {Harwood}}, \bibinfo {author} {\bibfnamefont {S.}~\bibnamefont {Vezzoli}}, \bibinfo {author} {\bibfnamefont {T.~V.}\ \bibnamefont {Raziman}}, \bibinfo {author} {\bibfnamefont {C.}~\bibnamefont {Hooper}}, \bibinfo {author} {\bibfnamefont {R.}~\bibnamefont {Tirole}}, \bibinfo {author} {\bibfnamefont {F.}~\bibnamefont {Wu}}, \bibinfo {author} {\bibfnamefont {S.~A.}\ \bibnamefont {Maier}}, \bibinfo {author} {\bibfnamefont {J.~B.}\ \bibnamefont {Pendry}}, \bibinfo {author} {\bibfnamefont {S.~A.~R.}\ \bibnamefont {Horsley}},\ and\ \bibinfo {author} {\bibfnamefont {R.}~\bibnamefont {Sapienza}},\ }\bibfield  {title} {\bibinfo {title} {Space-time optical diffraction from synthetic motion},\ }\href {https://doi.org/10.1038/s41467-025-60159-9} {\bibfield  {journal} {\bibinfo  {journal} {Nature Communications}\ }\textbf {\bibinfo {volume} {16}},\ \bibinfo {pages} {5147} (\bibinfo {year} {2025})}\BibitemShut {NoStop}%
\bibitem [{\citenamefont {Asgari}\ \emph {et~al.}(2024)\citenamefont {Asgari}, \citenamefont {Garg}, \citenamefont {Wang}, \citenamefont {Mirmoosa}, \citenamefont {Rockstuhl},\ and\ \citenamefont {Asadchy}}]{Asgari2024}%
  \BibitemOpen
  \bibfield  {author} {\bibinfo {author} {\bibfnamefont {M.~M.}\ \bibnamefont {Asgari}}, \bibinfo {author} {\bibfnamefont {P.}~\bibnamefont {Garg}}, \bibinfo {author} {\bibfnamefont {X.}~\bibnamefont {Wang}}, \bibinfo {author} {\bibfnamefont {M.~S.}\ \bibnamefont {Mirmoosa}}, \bibinfo {author} {\bibfnamefont {C.}~\bibnamefont {Rockstuhl}},\ and\ \bibinfo {author} {\bibfnamefont {V.}~\bibnamefont {Asadchy}},\ }\bibfield  {title} {\bibinfo {title} {Theory and applications of photonic time crystals: a tutorial},\ }\href {https://doi.org/10.1364/AOP.525163} {\bibfield  {journal} {\bibinfo  {journal} {Adv. Opt. Photon.}\ }\textbf {\bibinfo {volume} {16}},\ \bibinfo {pages} {958} (\bibinfo {year} {2024})}\BibitemShut {NoStop}%
\bibitem [{\citenamefont {Lustig}\ \emph {et~al.}(2023)\citenamefont {Lustig}, \citenamefont {Segal}, \citenamefont {Saha}, \citenamefont {Fruhling}, \citenamefont {Shalaev}, \citenamefont {Boltasseva},\ and\ \citenamefont {Segev}}]{Lustig2023}%
  \BibitemOpen
  \bibfield  {author} {\bibinfo {author} {\bibfnamefont {E.}~\bibnamefont {Lustig}}, \bibinfo {author} {\bibfnamefont {O.}~\bibnamefont {Segal}}, \bibinfo {author} {\bibfnamefont {S.}~\bibnamefont {Saha}}, \bibinfo {author} {\bibfnamefont {C.}~\bibnamefont {Fruhling}}, \bibinfo {author} {\bibfnamefont {V.~M.}\ \bibnamefont {Shalaev}}, \bibinfo {author} {\bibfnamefont {A.}~\bibnamefont {Boltasseva}},\ and\ \bibinfo {author} {\bibfnamefont {M.}~\bibnamefont {Segev}},\ }\bibfield  {title} {\bibinfo {title} {Photonic time-crystals - fundamental concepts [invited]},\ }\href {https://doi.org/10.1364/OE.479367} {\bibfield  {journal} {\bibinfo  {journal} {Opt. Express}\ }\textbf {\bibinfo {volume} {31}},\ \bibinfo {pages} {9165} (\bibinfo {year} {2023})}\BibitemShut {NoStop}%
\bibitem [{\citenamefont {He}\ \emph {et~al.}(2023)\citenamefont {He}, \citenamefont {Zhang}, \citenamefont {Qi}, \citenamefont {Bo},\ and\ \citenamefont {Li}}]{He2023}%
  \BibitemOpen
  \bibfield  {author} {\bibinfo {author} {\bibfnamefont {H.}~\bibnamefont {He}}, \bibinfo {author} {\bibfnamefont {S.}~\bibnamefont {Zhang}}, \bibinfo {author} {\bibfnamefont {J.}~\bibnamefont {Qi}}, \bibinfo {author} {\bibfnamefont {F.}~\bibnamefont {Bo}},\ and\ \bibinfo {author} {\bibfnamefont {H.}~\bibnamefont {Li}},\ }\bibfield  {title} {\bibinfo {title} {Faraday rotation in nonreciprocal photonic time-crystals},\ }\href {https://doi.org/10.1063/5.0131818} {\bibfield  {journal} {\bibinfo  {journal} {Applied Physics Letters}\ }\textbf {\bibinfo {volume} {122}},\ \bibinfo {pages} {051703} (\bibinfo {year} {2023})},\ \Eprint {https://arxiv.org/abs/https://pubs.aip.org/aip/apl/article-pdf/doi/10.1063/5.0131818/16733058/051703\_1\_online.pdf} {https://pubs.aip.org/aip/apl/article-pdf/doi/10.1063/5.0131818/16733058/051703\_1\_online.pdf} \BibitemShut {NoStop}%
\bibitem [{\citenamefont {Gaxiola-Luna}\ and\ \citenamefont {Halevi}(2023)}]{GaxiolaLuna2023}%
  \BibitemOpen
  \bibfield  {author} {\bibinfo {author} {\bibfnamefont {J.~G.}\ \bibnamefont {Gaxiola-Luna}}\ and\ \bibinfo {author} {\bibfnamefont {P.}~\bibnamefont {Halevi}},\ }\bibfield  {title} {\bibinfo {title} {Growing fields in a temporal photonic (time) crystal with a square profile of the permittivity $\epsilon(t)$},\ }\href {https://doi.org/10.1063/5.0132906} {\bibfield  {journal} {\bibinfo  {journal} {Applied Physics Letters}\ }\textbf {\bibinfo {volume} {122}},\ \bibinfo {pages} {011702} (\bibinfo {year} {2023})},\ \Eprint {https://arxiv.org/abs/https://pubs.aip.org/aip/apl/article-pdf/doi/10.1063/5.0132906/16746729/011702\_1\_online.pdf} {https://pubs.aip.org/aip/apl/article-pdf/doi/10.1063/5.0132906/16746729/011702\_1\_online.pdf} \BibitemShut {NoStop}%
\bibitem [{\citenamefont {Trainiti}\ \emph {et~al.}(2019)\citenamefont {Trainiti}, \citenamefont {Xia}, \citenamefont {Marconi}, \citenamefont {Cazzulani}, \citenamefont {Erturk},\ and\ \citenamefont {Ruzzene}}]{Trainiti2019}%
  \BibitemOpen
  \bibfield  {author} {\bibinfo {author} {\bibfnamefont {G.}~\bibnamefont {Trainiti}}, \bibinfo {author} {\bibfnamefont {Y.}~\bibnamefont {Xia}}, \bibinfo {author} {\bibfnamefont {J.}~\bibnamefont {Marconi}}, \bibinfo {author} {\bibfnamefont {G.}~\bibnamefont {Cazzulani}}, \bibinfo {author} {\bibfnamefont {A.}~\bibnamefont {Erturk}},\ and\ \bibinfo {author} {\bibfnamefont {M.}~\bibnamefont {Ruzzene}},\ }\bibfield  {title} {\bibinfo {title} {Time-periodic stiffness modulation in elastic metamaterials for selective wave filtering: Theory and experiment},\ }\href {https://doi.org/10.1103/PhysRevLett.122.124301} {\bibfield  {journal} {\bibinfo  {journal} {Phys. Rev. Lett.}\ }\textbf {\bibinfo {volume} {122}},\ \bibinfo {pages} {124301} (\bibinfo {year} {2019})}\BibitemShut {NoStop}%
\bibitem [{\citenamefont {Hooper}\ \emph {et~al.}(2025)\citenamefont {Hooper}, \citenamefont {Capers}, \citenamefont {Hooper},\ and\ \citenamefont {Horsley}}]{Hooper2025}%
  \BibitemOpen
  \bibfield  {author} {\bibinfo {author} {\bibfnamefont {C.~M.}\ \bibnamefont {Hooper}}, \bibinfo {author} {\bibfnamefont {J.~R.}\ \bibnamefont {Capers}}, \bibinfo {author} {\bibfnamefont {I.~R.}\ \bibnamefont {Hooper}},\ and\ \bibinfo {author} {\bibfnamefont {S.~A.~R.}\ \bibnamefont {Horsley}},\ }\bibfield  {title} {\bibinfo {title} {Symmetry-protected lossless modes in dispersive time-varying media},\ }\href {https://doi.org/10.1103/PhysRevA.111.033507} {\bibfield  {journal} {\bibinfo  {journal} {Phys. Rev. A}\ }\textbf {\bibinfo {volume} {111}},\ \bibinfo {pages} {033507} (\bibinfo {year} {2025})}\BibitemShut {NoStop}%
\bibitem [{\citenamefont {Galiffi}\ \emph {et~al.}(2024)\citenamefont {Galiffi}, \citenamefont {Harwood}, \citenamefont {Vezzoli}, \citenamefont {Tirole}, \citenamefont {Alù},\ and\ \citenamefont {Sapienza}}]{Galiffi2024}%
  \BibitemOpen
  \bibfield  {author} {\bibinfo {author} {\bibfnamefont {E.}~\bibnamefont {Galiffi}}, \bibinfo {author} {\bibfnamefont {A.~C.}\ \bibnamefont {Harwood}}, \bibinfo {author} {\bibfnamefont {S.}~\bibnamefont {Vezzoli}}, \bibinfo {author} {\bibfnamefont {R.}~\bibnamefont {Tirole}}, \bibinfo {author} {\bibfnamefont {A.}~\bibnamefont {Alù}},\ and\ \bibinfo {author} {\bibfnamefont {R.}~\bibnamefont {Sapienza}},\ }\href {https://arxiv.org/abs/2410.16426} {\bibinfo {title} {Optical coherent perfect absorption and amplification in a time-varying medium}} (\bibinfo {year} {2024}),\ \Eprint {https://arxiv.org/abs/2410.16426} {arXiv:2410.16426 [physics.optics]} \BibitemShut {NoStop}%
\bibitem [{\citenamefont {Hendry}\ \emph {et~al.}(2025)\citenamefont {Hendry}, \citenamefont {Hooper}, \citenamefont {Wardley},\ and\ \citenamefont {Horsley}}]{Hendry2025}%
  \BibitemOpen
  \bibfield  {author} {\bibinfo {author} {\bibfnamefont {E.}~\bibnamefont {Hendry}}, \bibinfo {author} {\bibfnamefont {C.~M.}\ \bibnamefont {Hooper}}, \bibinfo {author} {\bibfnamefont {W.~P.}\ \bibnamefont {Wardley}},\ and\ \bibinfo {author} {\bibfnamefont {S.~A.~R.}\ \bibnamefont {Horsley}},\ }\href {https://arxiv.org/abs/2507.03491} {\bibinfo {title} {Effects due to generation of negative frequencies during temporal diffraction}} (\bibinfo {year} {2025}),\ \Eprint {https://arxiv.org/abs/2507.03491} {arXiv:2507.03491 [physics.optics]} \BibitemShut {NoStop}%
\bibitem [{\citenamefont {Wang}\ \emph {et~al.}(2023)\citenamefont {Wang}, \citenamefont {Mirmoosa}, \citenamefont {Asadchy}, \citenamefont {Rockstuhl}, \citenamefont {Fan},\ and\ \citenamefont {Tretyakov}}]{Wang2023}%
  \BibitemOpen
  \bibfield  {author} {\bibinfo {author} {\bibfnamefont {X.}~\bibnamefont {Wang}}, \bibinfo {author} {\bibfnamefont {M.~S.}\ \bibnamefont {Mirmoosa}}, \bibinfo {author} {\bibfnamefont {V.~S.}\ \bibnamefont {Asadchy}}, \bibinfo {author} {\bibfnamefont {C.}~\bibnamefont {Rockstuhl}}, \bibinfo {author} {\bibfnamefont {S.}~\bibnamefont {Fan}},\ and\ \bibinfo {author} {\bibfnamefont {S.~A.}\ \bibnamefont {Tretyakov}},\ }\bibfield  {title} {\bibinfo {title} {Metasurface-based realization of photonic time crystals},\ }\href {https://doi.org/10.1126/sciadv.adg7541} {\bibfield  {journal} {\bibinfo  {journal} {Science Advances}\ }\textbf {\bibinfo {volume} {9}},\ \bibinfo {pages} {eadg7541} (\bibinfo {year} {2023})},\ \Eprint {https://arxiv.org/abs/https://www.science.org/doi/pdf/10.1126/sciadv.adg7541} {https://www.science.org/doi/pdf/10.1126/sciadv.adg7541} \BibitemShut {NoStop}%
\bibitem [{\citenamefont {Vertiz-Conde}\ \emph {et~al.}(2025)\citenamefont {Vertiz-Conde}, \citenamefont {{n}igo Liberal},\ and\ \citenamefont {V\'{a}zquez-Lozano}}]{VertizConde2025}%
  \BibitemOpen
  \bibfield  {author} {\bibinfo {author} {\bibfnamefont {A.}~\bibnamefont {Vertiz-Conde}}, \bibinfo {author} {\bibfnamefont {I.}~\bibnamefont {{n}igo Liberal}},\ and\ \bibinfo {author} {\bibfnamefont {J.~E.}\ \bibnamefont {V\'{a}zquez-Lozano}},\ }\bibfield  {title} {\bibinfo {title} {Dispersion effects in thermal emission from temporal metamaterials: high-frequency cutoffs},\ }\href {https://doi.org/10.1364/OL.545236} {\bibfield  {journal} {\bibinfo  {journal} {Opt. Lett.}\ }\textbf {\bibinfo {volume} {50}},\ \bibinfo {pages} {1097} (\bibinfo {year} {2025})}\BibitemShut {NoStop}%
\bibitem [{\citenamefont {V{\'a}zquez-Lozano}\ and\ \citenamefont {Liberal}(2023)}]{VazquezLozano2023}%
  \BibitemOpen
  \bibfield  {author} {\bibinfo {author} {\bibfnamefont {J.~E.}\ \bibnamefont {V{\'a}zquez-Lozano}}\ and\ \bibinfo {author} {\bibfnamefont {I.}~\bibnamefont {Liberal}},\ }\bibfield  {title} {\bibinfo {title} {Incandescent temporal metamaterials},\ }\href {https://doi.org/10.1038/s41467-023-40281-2} {\bibfield  {journal} {\bibinfo  {journal} {Nature Communications}\ }\textbf {\bibinfo {volume} {14}},\ \bibinfo {pages} {4606} (\bibinfo {year} {2023})}\BibitemShut {NoStop}%
\bibitem [{\citenamefont {Horsley}\ and\ \citenamefont {Pendry}(2023)}]{Horsley2023BlackHole}%
  \BibitemOpen
  \bibfield  {author} {\bibinfo {author} {\bibfnamefont {S.~A.~R.}\ \bibnamefont {Horsley}}\ and\ \bibinfo {author} {\bibfnamefont {J.~B.}\ \bibnamefont {Pendry}},\ }\bibfield  {title} {\bibinfo {title} {Quantum electrodynamics of time-varying gratings},\ }\href {https://doi.org/10.1073/pnas.2302652120} {\bibfield  {journal} {\bibinfo  {journal} {Proceedings of the National Academy of Sciences}\ }\textbf {\bibinfo {volume} {120}},\ \bibinfo {pages} {e2302652120} (\bibinfo {year} {2023})},\ \Eprint {https://arxiv.org/abs/https://www.pnas.org/doi/pdf/10.1073/pnas.2302652120} {https://www.pnas.org/doi/pdf/10.1073/pnas.2302652120} \BibitemShut {NoStop}%
\bibitem [{\citenamefont {Horsley}\ and\ \citenamefont {Pendry}(2024)}]{Horsley2024}%
  \BibitemOpen
  \bibfield  {author} {\bibinfo {author} {\bibfnamefont {S.~A.~R.}\ \bibnamefont {Horsley}}\ and\ \bibinfo {author} {\bibfnamefont {J.~B.}\ \bibnamefont {Pendry}},\ }\bibfield  {title} {\bibinfo {title} {Traveling wave amplification in stationary gratings},\ }\href {https://doi.org/10.1103/PhysRevLett.133.156903} {\bibfield  {journal} {\bibinfo  {journal} {Phys. Rev. Lett.}\ }\textbf {\bibinfo {volume} {133}},\ \bibinfo {pages} {156903} (\bibinfo {year} {2024})}\BibitemShut {NoStop}%
\bibitem [{\citenamefont {Horsley}\ \emph {et~al.}(2023)\citenamefont {Horsley}, \citenamefont {Galiffi},\ and\ \citenamefont {Wang}}]{Horsley2023Operators}%
  \BibitemOpen
  \bibfield  {author} {\bibinfo {author} {\bibfnamefont {S.~A.~R.}\ \bibnamefont {Horsley}}, \bibinfo {author} {\bibfnamefont {E.}~\bibnamefont {Galiffi}},\ and\ \bibinfo {author} {\bibfnamefont {Y.-T.}\ \bibnamefont {Wang}},\ }\bibfield  {title} {\bibinfo {title} {Eigenpulses of dispersive time-varying media},\ }\href {https://doi.org/10.1103/PhysRevLett.130.203803} {\bibfield  {journal} {\bibinfo  {journal} {Phys. Rev. Lett.}\ }\textbf {\bibinfo {volume} {130}},\ \bibinfo {pages} {203803} (\bibinfo {year} {2023})}\BibitemShut {NoStop}%
\bibitem [{\citenamefont {Tomadin}\ \emph {et~al.}(2018)\citenamefont {Tomadin}, \citenamefont {Hornett}, \citenamefont {Wang}, \citenamefont {Alexeev}, \citenamefont {Candini}, \citenamefont {Coletti}, \citenamefont {Turchinovich}, \citenamefont {Kläui}, \citenamefont {Bonn}, \citenamefont {Koppens}, \citenamefont {Hendry}, \citenamefont {Polini},\ and\ \citenamefont {Tielrooij}}]{Tomadin2018}%
  \BibitemOpen
  \bibfield  {author} {\bibinfo {author} {\bibfnamefont {A.}~\bibnamefont {Tomadin}}, \bibinfo {author} {\bibfnamefont {S.~M.}\ \bibnamefont {Hornett}}, \bibinfo {author} {\bibfnamefont {H.~I.}\ \bibnamefont {Wang}}, \bibinfo {author} {\bibfnamefont {E.~M.}\ \bibnamefont {Alexeev}}, \bibinfo {author} {\bibfnamefont {A.}~\bibnamefont {Candini}}, \bibinfo {author} {\bibfnamefont {C.}~\bibnamefont {Coletti}}, \bibinfo {author} {\bibfnamefont {D.}~\bibnamefont {Turchinovich}}, \bibinfo {author} {\bibfnamefont {M.}~\bibnamefont {Kläui}}, \bibinfo {author} {\bibfnamefont {M.}~\bibnamefont {Bonn}}, \bibinfo {author} {\bibfnamefont {F.~H.~L.}\ \bibnamefont {Koppens}}, \bibinfo {author} {\bibfnamefont {E.}~\bibnamefont {Hendry}}, \bibinfo {author} {\bibfnamefont {M.}~\bibnamefont {Polini}},\ and\ \bibinfo {author} {\bibfnamefont {K.-J.}\ \bibnamefont {Tielrooij}},\ }\bibfield  {title} {\bibinfo {title} {The ultrafast dynamics and conductivity of photoexcited graphene at different fermi energies},\ }\href
  {https://doi.org/10.1126/sciadv.aar5313} {\bibfield  {journal} {\bibinfo  {journal} {Science Advances}\ }\textbf {\bibinfo {volume} {4}},\ \bibinfo {pages} {eaar5313} (\bibinfo {year} {2018})},\ \Eprint {https://arxiv.org/abs/https://www.science.org/doi/pdf/10.1126/sciadv.aar5313} {https://www.science.org/doi/pdf/10.1126/sciadv.aar5313} \BibitemShut {NoStop}%
\bibitem [{\citenamefont {Moussa}\ \emph {et~al.}(2023)\citenamefont {Moussa}, \citenamefont {Xu}, \citenamefont {Yin}, \citenamefont {Galiffi}, \citenamefont {Ra'di},\ and\ \citenamefont {Al{\`u}}}]{Moussa2023}%
  \BibitemOpen
  \bibfield  {author} {\bibinfo {author} {\bibfnamefont {H.}~\bibnamefont {Moussa}}, \bibinfo {author} {\bibfnamefont {G.}~\bibnamefont {Xu}}, \bibinfo {author} {\bibfnamefont {S.}~\bibnamefont {Yin}}, \bibinfo {author} {\bibfnamefont {E.}~\bibnamefont {Galiffi}}, \bibinfo {author} {\bibfnamefont {Y.}~\bibnamefont {Ra'di}},\ and\ \bibinfo {author} {\bibfnamefont {A.}~\bibnamefont {Al{\`u}}},\ }\bibfield  {title} {\bibinfo {title} {Observation of temporal reflection and broadband frequency translation at photonic time interfaces},\ }\href {https://doi.org/10.1038/s41567-023-01975-y} {\bibfield  {journal} {\bibinfo  {journal} {Nature Physics}\ }\textbf {\bibinfo {volume} {19}},\ \bibinfo {pages} {863} (\bibinfo {year} {2023})}\BibitemShut {NoStop}%
\bibitem [{\citenamefont {Wang}\ \emph {et~al.}(2025)\citenamefont {Wang}, \citenamefont {Garg}, \citenamefont {Mirmoosa}, \citenamefont {Lamprianidis}, \citenamefont {Rockstuhl},\ and\ \citenamefont {Asadchy}}]{Wang2025}%
  \BibitemOpen
  \bibfield  {author} {\bibinfo {author} {\bibfnamefont {X.}~\bibnamefont {Wang}}, \bibinfo {author} {\bibfnamefont {P.}~\bibnamefont {Garg}}, \bibinfo {author} {\bibfnamefont {M.~S.}\ \bibnamefont {Mirmoosa}}, \bibinfo {author} {\bibfnamefont {A.~G.}\ \bibnamefont {Lamprianidis}}, \bibinfo {author} {\bibfnamefont {C.}~\bibnamefont {Rockstuhl}},\ and\ \bibinfo {author} {\bibfnamefont {V.~S.}\ \bibnamefont {Asadchy}},\ }\bibfield  {title} {\bibinfo {title} {Expanding momentum bandgaps in photonic time crystals through resonances},\ }\href {https://doi.org/10.1038/s41566-024-01563-3} {\bibfield  {journal} {\bibinfo  {journal} {Nature Photonics}\ }\textbf {\bibinfo {volume} {19}},\ \bibinfo {pages} {149} (\bibinfo {year} {2025})}\BibitemShut {NoStop}%
\bibitem [{\citenamefont {Benisty}\ \emph {et~al.}(2022)\citenamefont {Benisty}, \citenamefont {Greffet},\ and\ \citenamefont {Lalanne}}]{Benisty2022}%
  \BibitemOpen
  \bibfield  {author} {\bibinfo {author} {\bibfnamefont {H.}~\bibnamefont {Benisty}}, \bibinfo {author} {\bibfnamefont {J.-J.}\ \bibnamefont {Greffet}},\ and\ \bibinfo {author} {\bibfnamefont {P.}~\bibnamefont {Lalanne}},\ }\bibfield  {title} {\bibinfo {title} {159basics of resonators and cavities},\ }in\ \href {https://doi.org/10.1093/oso/9780198786139.003.0007} {\emph {\bibinfo {booktitle} {Introduction to Nanophotonics}}}\ (\bibinfo  {publisher} {Oxford University Press},\ \bibinfo {year} {2022})\ \Eprint {https://arxiv.org/abs/https://academic.oup.com/book/0/chapter/374111973/chapter-pdf/50294469/oso-9780198786139-chapter-7.pdf} {https://academic.oup.com/book/0/chapter/374111973/chapter-pdf/50294469/oso-9780198786139-chapter-7.pdf} \BibitemShut {NoStop}%
\bibitem [{\citenamefont {Kristensen}\ and\ \citenamefont {Hughes}(2014)}]{Kristensen2014}%
  \BibitemOpen
  \bibfield  {author} {\bibinfo {author} {\bibfnamefont {P.~T.}\ \bibnamefont {Kristensen}}\ and\ \bibinfo {author} {\bibfnamefont {S.}~\bibnamefont {Hughes}},\ }\bibfield  {title} {\bibinfo {title} {Modes and mode volumes of leaky optical cavities and plasmonic nanoresonators},\ }\href {https://doi.org/10.1021/ph400114e} {\bibfield  {journal} {\bibinfo  {journal} {ACS Photonics}\ }\textbf {\bibinfo {volume} {1}},\ \bibinfo {pages} {2} (\bibinfo {year} {2014})}\BibitemShut {NoStop}%
\bibitem [{\citenamefont {Valero}\ \emph {et~al.}(2025)\citenamefont {Valero}, \citenamefont {Gladyshev}, \citenamefont {Globosits}, \citenamefont {Rotter}, \citenamefont {Muljarov},\ and\ \citenamefont {Weiss}}]{Valero2025}%
  \BibitemOpen
  \bibfield  {author} {\bibinfo {author} {\bibfnamefont {A.~C.}\ \bibnamefont {Valero}}, \bibinfo {author} {\bibfnamefont {S.}~\bibnamefont {Gladyshev}}, \bibinfo {author} {\bibfnamefont {D.}~\bibnamefont {Globosits}}, \bibinfo {author} {\bibfnamefont {S.}~\bibnamefont {Rotter}}, \bibinfo {author} {\bibfnamefont {E.~A.}\ \bibnamefont {Muljarov}},\ and\ \bibinfo {author} {\bibfnamefont {T.}~\bibnamefont {Weiss}},\ }\href {https://arxiv.org/abs/2506.01472} {\bibinfo {title} {Resonant states of structured photonic time crystals}} (\bibinfo {year} {2025}),\ \Eprint {https://arxiv.org/abs/2506.01472} {arXiv:2506.01472 [physics.optics]} \BibitemShut {NoStop}%
\bibitem [{\citenamefont {Vial}\ and\ \citenamefont {Craster}(2025)}]{Vial2025}%
  \BibitemOpen
  \bibfield  {author} {\bibinfo {author} {\bibfnamefont {B.}~\bibnamefont {Vial}}\ and\ \bibinfo {author} {\bibfnamefont {R.~V.}\ \bibnamefont {Craster}},\ }\href {https://arxiv.org/abs/2507.02784} {\bibinfo {title} {Quasinormal modes of floquet media slabs}} (\bibinfo {year} {2025}),\ \Eprint {https://arxiv.org/abs/2507.02784} {arXiv:2507.02784 [physics.optics]} \BibitemShut {NoStop}%
\bibitem [{\citenamefont {Sauvan}\ \emph {et~al.}(2022)\citenamefont {Sauvan}, \citenamefont {Wu}, \citenamefont {Zarouf}, \citenamefont {Muljarov},\ and\ \citenamefont {Lalanne}}]{Sauvan2022}%
  \BibitemOpen
  \bibfield  {author} {\bibinfo {author} {\bibfnamefont {C.}~\bibnamefont {Sauvan}}, \bibinfo {author} {\bibfnamefont {T.}~\bibnamefont {Wu}}, \bibinfo {author} {\bibfnamefont {R.}~\bibnamefont {Zarouf}}, \bibinfo {author} {\bibfnamefont {E.~A.}\ \bibnamefont {Muljarov}},\ and\ \bibinfo {author} {\bibfnamefont {P.}~\bibnamefont {Lalanne}},\ }\bibfield  {title} {\bibinfo {title} {Normalization, orthogonality, and completeness of quasinormal modes of open systems: the case of electromagnetism [invited]},\ }\href {https://doi.org/10.1364/OE.443656} {\bibfield  {journal} {\bibinfo  {journal} {Opt. Express}\ }\textbf {\bibinfo {volume} {30}},\ \bibinfo {pages} {6846} (\bibinfo {year} {2022})}\BibitemShut {NoStop}%
\bibitem [{\citenamefont {Lyubarov}\ \emph {et~al.}(2022)\citenamefont {Lyubarov}, \citenamefont {Lumer}, \citenamefont {Dikopoltsev}, \citenamefont {Lustig}, \citenamefont {Sharabi},\ and\ \citenamefont {Segev}}]{Lyubarov2022}%
  \BibitemOpen
  \bibfield  {author} {\bibinfo {author} {\bibfnamefont {M.}~\bibnamefont {Lyubarov}}, \bibinfo {author} {\bibfnamefont {Y.}~\bibnamefont {Lumer}}, \bibinfo {author} {\bibfnamefont {A.}~\bibnamefont {Dikopoltsev}}, \bibinfo {author} {\bibfnamefont {E.}~\bibnamefont {Lustig}}, \bibinfo {author} {\bibfnamefont {Y.}~\bibnamefont {Sharabi}},\ and\ \bibinfo {author} {\bibfnamefont {M.}~\bibnamefont {Segev}},\ }\bibfield  {title} {\bibinfo {title} {Amplified emission and lasing in photonic time crystals},\ }\href {https://doi.org/10.1126/science.abo3324} {\bibfield  {journal} {\bibinfo  {journal} {Science}\ }\textbf {\bibinfo {volume} {377}},\ \bibinfo {pages} {425} (\bibinfo {year} {2022})},\ \Eprint {https://arxiv.org/abs/https://www.science.org/doi/pdf/10.1126/science.abo3324} {https://www.science.org/doi/pdf/10.1126/science.abo3324} \BibitemShut {NoStop}%
\bibitem [{\citenamefont {Bender}\ and\ \citenamefont {Boettcher}(1998)}]{Bender1998}%
  \BibitemOpen
  \bibfield  {author} {\bibinfo {author} {\bibfnamefont {C.~M.}\ \bibnamefont {Bender}}\ and\ \bibinfo {author} {\bibfnamefont {S.}~\bibnamefont {Boettcher}},\ }\bibfield  {title} {\bibinfo {title} {Real spectra in non-hermitian hamiltonians having $\mathcal{PT}$ symmetry},\ }\href {https://doi.org/10.1103/PhysRevLett.80.5243} {\bibfield  {journal} {\bibinfo  {journal} {Phys. Rev. Lett.}\ }\textbf {\bibinfo {volume} {80}},\ \bibinfo {pages} {5243} (\bibinfo {year} {1998})}\BibitemShut {NoStop}%
\bibitem [{\citenamefont {Renardy}\ and\ \citenamefont {Rogers}(2004)}]{Renardy2004}%
  \BibitemOpen
  \bibfield  {author} {\bibinfo {author} {\bibfnamefont {M.}~\bibnamefont {Renardy}}\ and\ \bibinfo {author} {\bibfnamefont {R.~C.}\ \bibnamefont {Rogers}},\ }\bibinfo {title} {Operator theory},\ in\ \href {https://doi.org/10.1007/0-387-21687-1_8} {\emph {\bibinfo {booktitle} {An Introduction to Partial Differential Equations}}}\ (\bibinfo  {publisher} {Springer New York},\ \bibinfo {address} {New York, NY},\ \bibinfo {year} {2004})\ pp.\ \bibinfo {pages} {228--282}\BibitemShut {NoStop}%
\bibitem [{\citenamefont {Kiorpelidis}\ \emph {et~al.}(2024)\citenamefont {Kiorpelidis}, \citenamefont {Diakonos}, \citenamefont {Theocharis},\ and\ \citenamefont {Pagneux}}]{Kiorpelidis2024}%
  \BibitemOpen
  \bibfield  {author} {\bibinfo {author} {\bibfnamefont {I.}~\bibnamefont {Kiorpelidis}}, \bibinfo {author} {\bibfnamefont {F.~K.}\ \bibnamefont {Diakonos}}, \bibinfo {author} {\bibfnamefont {G.}~\bibnamefont {Theocharis}},\ and\ \bibinfo {author} {\bibfnamefont {V.}~\bibnamefont {Pagneux}},\ }\bibfield  {title} {\bibinfo {title} {Transient amplification in stable floquet media},\ }\href {https://doi.org/10.1103/PhysRevB.110.134315} {\bibfield  {journal} {\bibinfo  {journal} {Phys. Rev. B}\ }\textbf {\bibinfo {volume} {110}},\ \bibinfo {pages} {134315} (\bibinfo {year} {2024})}\BibitemShut {NoStop}%
\bibitem [{\citenamefont {Wang}(2013)}]{wang20132}%
  \BibitemOpen
  \bibfield  {author} {\bibinfo {author} {\bibfnamefont {Q.-h.}\ \bibnamefont {Wang}},\ }\bibfield  {title} {\bibinfo {title} {2$\times$ 2 pt-symmetric matrices and their applications},\ }\href@noop {} {\bibfield  {journal} {\bibinfo  {journal} {Philosophical Transactions of the Royal Society A: Mathematical, Physical and Engineering Sciences}\ }\textbf {\bibinfo {volume} {371}},\ \bibinfo {pages} {20120045} (\bibinfo {year} {2013})}\BibitemShut {NoStop}%
\bibitem [{\citenamefont {Hanson}\ and\ \citenamefont {Yakovlev}(2002)}]{Hanson2002}%
  \BibitemOpen
  \bibfield  {author} {\bibinfo {author} {\bibfnamefont {G.~W.}\ \bibnamefont {Hanson}}\ and\ \bibinfo {author} {\bibfnamefont {A.~B.}\ \bibnamefont {Yakovlev}},\ }\bibinfo {title} {Introductory linear operator theory},\ in\ \href {https://doi.org/10.1007/978-1-4757-3679-3_3} {\emph {\bibinfo {booktitle} {Operator Theory for Electromagnetics: An Introduction}}}\ (\bibinfo  {publisher} {Springer New York},\ \bibinfo {address} {New York, NY},\ \bibinfo {year} {2002})\ pp.\ \bibinfo {pages} {129--217}\BibitemShut {NoStop}%
\bibitem [{\citenamefont {Zweck}\ \emph {et~al.}(2025)\citenamefont {Zweck}, \citenamefont {Latushkin},\ and\ \citenamefont {Gallo}}]{Zweck2025}%
  \BibitemOpen
  \bibfield  {author} {\bibinfo {author} {\bibfnamefont {J.}~\bibnamefont {Zweck}}, \bibinfo {author} {\bibfnamefont {Y.}~\bibnamefont {Latushkin}},\ and\ \bibinfo {author} {\bibfnamefont {E.}~\bibnamefont {Gallo}},\ }\bibfield  {title} {\bibinfo {title} {A regularity condition under which integral operators with operator-valued kernels are trace class},\ }\href {https://doi.org/10.1007/s40590-025-00718-8} {\bibfield  {journal} {\bibinfo  {journal} {Bolet{\'i}n de la Sociedad Matem{\'a}tica Mexicana}\ }\textbf {\bibinfo {volume} {31}},\ \bibinfo {pages} {38} (\bibinfo {year} {2025})}\BibitemShut {NoStop}%
\bibitem [{\citenamefont {Steinberg}(1968)}]{Steinberg1968}%
  \BibitemOpen
  \bibfield  {author} {\bibinfo {author} {\bibfnamefont {S.}~\bibnamefont {Steinberg}},\ }\bibfield  {title} {\bibinfo {title} {Meromorphic families of compact operators},\ }\href {https://doi.org/10.1007/BF00251419} {\bibfield  {journal} {\bibinfo  {journal} {Archive for Rational Mechanics and Analysis}\ }\textbf {\bibinfo {volume} {31}},\ \bibinfo {pages} {372} (\bibinfo {year} {1968})}\BibitemShut {NoStop}%
\end{thebibliography}%

\appendix

\section{\label{app:FQNMEigenvalueProblem}FQNMs as a (Non-)Linear Eigenvalue Problem}
It is worth briefly justifying the sense in which the FQNM problem as derived from Equation (\ref{eq:TimeDomain1DWaveEquation}) (with outgoing boundary conditions) is related to an eigenvalue problem. To this end, we consider the problem

\begin{equation}
    \left( \partial_{t} - \begin{pmatrix}
    L_{\rm WW} & L_{\rm WM} \\
    L_{\rm MW} & L_{\rm MM}
    \end{pmatrix} \right)\left( \begin{array}{r}
    \psi_{\rm W} \\
    \psi_{\rm M}
    \end{array} \right) = 0,
    \label{eq:GeneralSplitTimeDomainFQNMCondition}
\end{equation}

\noindent where \(\psi_{\rm W}\) and \(\psi_{\rm M}\) are real fields associated with the wave and material, respectively. Throughout this section, explicit dependence on space and time will be taken as implicit. On their own, these fields would evolve with \(\partial_{t}\psi_{\rm Q} = L_{\rm QQ}\psi_{\rm Q}\) for \(\rm Q \in \left\{ W,M \right\}\), where \(L_{\rm WW}\) is assumed to act on the space of outgoing waves \(\psi_{\rm W}\). For our problem, these fields are coupled by the operators \(L_{\rm WM}\) and \(L_{\rm MW}\), respectively describing coupling from the material to the wave, and vice versa. Throughout this work, we will frequently assume that we are dealing with light interacting with some dispersive material, such that \(L_{\rm WW}\) contains no time-variation, with any time-variations contained either in the evolution of the material field, or in the coupling to/from it. More compactly, this can be written as

\begin{equation}
    \left( \partial_{t} - L \right)\Psi = 0,
    \label{eq:SimpleTimeDomainFQNMEquation}
\end{equation}

\noindent where

\begin{equation}
    \Psi = \left( \begin{array}{r}
    \psi_{\rm W} \\
    \psi_{\rm M}
    \end{array} \right),
    \label{eq:SplitFieldsDefinition}
\end{equation}

and

\begin{equation}
    L=\begin{pmatrix}
    L_{\rm WW} & L_{\rm WM} \\
    L_{\rm MW} & L_{\rm MM}
    \end{pmatrix}.
    \label{eq:TimeDomainOperatorComponents}
\end{equation}

We briefly note that, since \(\Psi\) consists entirely of real fields, \(L^{*} = L\). Finally, we introduce the periodicity constraint of time-crystals, that \(L\left( t + \frac{2\pi}{\Omega} \right) = L(t)\), at which stage it also becomes convenient to introduce the Floquet ansatz for solutions:

\begin{equation}
    \Psi\left( t;\omega_{0} \right) = {\rm e}^{- \mathrm{i}\omega_{0}t}\overline{\Psi}\left( t;\omega_{0} \right),
    \label{eq:FloquetFieldDefinition}
\end{equation}

\noindent where \(\overline{\Psi}\left( t + \frac{2\pi}{\Omega};\omega_{0} \right) = \overline{\Psi}\left( t;\omega_{0} \right)\). Substituting this ansatz into Equation (\ref{eq:SimpleTimeDomainFQNMEquation}) gives

\begin{equation}
    \left( \partial_{t} - \mathrm{i}\omega_{0} - L \right)\overline{\Psi}\left( \omega_{0} \right) = 0.
    \label{eq:TimeDomainFloquetWaveEquation}
\end{equation}

However, the periodicity of \(L\) also implies that it can be expressed as the Fourier series,

\begin{equation}
    L(t) = \sum_{n = - \infty}^{\infty}{{\rm e}^{- \mathrm{i}n\Omega t}L_{n}},
    \label{eq:TimeDomainPeriodicOperatorDefinition}
\end{equation}

\noindent where \(\left\lbrack \partial_{t},L_{n} \right\rbrack = 0\), and \(L_{n} = L_{- n}^{*}\). With this, we then apply the method of \cite{Horsley2023Operators} to convert this problem to the frequency domain with the mappings \(\overline{\Psi}\left( t;\omega_{0} \right) \mapsto \widetilde{\Psi}\left( \omega;\omega_{0} \right)\), \(\partial_{t} \mapsto - \mathrm{i}\omega\), and \(t \mapsto - \mathrm{i}\partial_{\omega}\), giving,

\begin{equation}
    \widetilde{L} = \sum_{n = - \infty}^{\infty}{{\rm e}^{- n\Omega\partial_{\omega}}{\widetilde{L}}_{n}},
    \label{eq:FrequencyDomainPeriodicOperatorDefinition}
\end{equation}

\noindent where \(\left\lbrack {\widetilde{L}}_{n},\omega \right\rbrack = 0\), and \({\rm e}^{- n\Omega\partial_{\omega}}\) are powers of the discrete frequency translation operator,

\begin{equation}
    {\rm e}^{- \Omega\partial_{\omega}}\widetilde{\Psi}\left( \omega;\omega_{0} \right) = \widetilde{\Psi}\left( \omega - \Omega;\omega_{0} \right).
    \label{eq:TranslationOperatorDefinition}
\end{equation}

We are then left with the following linear eigenvalue problem in \(\omega_{0}\):

\begin{equation}
    \left( - \mathrm{i}\left( \omega + \omega_{0} \right) - \widetilde{L} \right)\widetilde{\Psi}\left( \omega;\omega_{0} \right) = 0.
    \label{eq:SimpleLinearFQNMFormula}
\end{equation}

Practically, however, most QNM calculations do not take close account of material degrees of freedom, instead working with wave equations similar in form to Equation (\ref{eq:TimeDomain1DWaveEquation}), where the material degrees of freedom have been eliminated and replaced with some general susceptibility-like operator encoding the material response. To this end, we re-expand Equation (\ref{eq:SimpleLinearFQNMFormula}) in terms of its wave and material components:

\begin{equation}
    \left( - \mathrm{i}\left( \omega + \omega_{0} \right) - \begin{pmatrix}
    {\widetilde{L}}_{\rm WW} & {\widetilde{L}}_{\rm WM} \\
    {\widetilde{L}}_{\rm MW} & {\widetilde{L}}_{\rm MM}
    \end{pmatrix} \right)\left( \begin{array}{r}
    {\widetilde{\psi}}_{\rm W}\left( \omega;\omega_{0} \right) \\
    {\widetilde{\psi}}_{\rm M}\left( \omega;\omega_{0} \right)
    \end{array} \right) = 0.
    \label{eq:FQNMCouplingEquation}
\end{equation}

Considering the lower equation, we find

\begin{equation}
    {\widetilde{\psi}}_{\rm M}\left( \omega;\omega_{0} \right) = \left( - \mathrm{i}\left( \omega + \omega_{0} \right) - {\widetilde{L}}_{\rm MM} \right)^{- 1}{\widetilde{L}}_{\rm MW}{\widetilde{\psi}}_{\rm W}\left( \omega;\omega_{0} \right).
    \label{eq:MaterialResponse}
\end{equation}

Thus, under the condition that we are away from any resonance of the decoupled material field (i.e., \(\left( - \mathrm{i}\left( \omega + \omega_{0} \right) - {\widetilde{L}}_{\rm MM} \right)^{- 1}\) is bounded for the \(\omega_{0}\) under consideration), it follows that the material fields can indeed be successfully eliminated, resulting in the nonlinear eigenvalue problem,

\begin{multline}
    \left( - \mathrm{i}\left( \omega + \omega_{0} \right) - {\widetilde{L}}_{\rm WW} \right){\widetilde{\psi}}_{\rm W}\left( \omega;\omega_{0} \right) =\\ {\widetilde{L}}_{\rm WM}\left( - \mathrm{i}\left( \omega + \omega_{0} \right) - {\widetilde{L}}_{\rm MM} \right)^{- 1}{\widetilde{L}}_{\rm MW}{\widetilde{\psi}}_{\rm W}\left( \omega;\omega_{0} \right),
    \label{eq:GeneralisedWaveEquationWithMaterial}
\end{multline}

\noindent defined away from \(\omega_{0}\) associated with material resonance. Here, the left-hand side corresponds to the free space wave equation, whilst the right-hand side corresponds to the material response normally encoded in the susceptibility.

\section{\label{app:FQNMFundamentals}Fundamental Properties of FQNMs and Their Trajectories}

Fredholm theory provides a number of useful guarantees regarding the distribution of solutions to nonlinear eigenvalue problems in the complex plane~\cite{Hanson2002, Renardy2004}, under the restriction that said generalised eigenvalue problem is written in terms of a compact operator. In this section, we demonstrate that these results can be applied to the FQNM eigenvalue problem, at least in the case of slabs in \(1\)-dimension, before conjecturing a generalisation to arbitrary dimensions.

The theorems we would like to apply refer to nonlinear eigenvalue problems of the form \(\left( 1 - A\left( \omega_{0} \right) \right)\Psi = 0\), where \(A\left( \omega_{0} \right)\) is a compact operator, varying holomorphically in \(\omega_{0}\) over some set of complex quasifrequencies. Since most operators associated with this problem (for instance the susceptibility operator) are holomorphic excluding a countably infinite set of points, the major difficulty is in phrasing our eigenvalue problem in terms of a compact \(A\left( \omega_{0} \right)\), thus providing a nonlinear eigenvalue problem in the Fredholm form above.

To this end, we consider the following rearrangement of Equation (\ref{eq:GeneralisedWaveEquationWithMaterial}):

\begin{equation}
    \left( 1 - G_{\rm WW}{\widetilde{L}}_{\rm WM}G_{\rm MM}{\widetilde{L}}_{\rm MW} \right){\widetilde{\psi}}_{\rm W} = 0,
    \label{eq:BornEigenvalues}
\end{equation}

\noindent where, for brevity, we let \(\left( - \mathrm{i}\left( \omega + \omega_{0} \right) - {\widetilde{L}}_{\rm QQ} \right)^{- 1} = G_{\rm QQ}\) for \(\rm Q \in \left\{ W,M \right\}\). This equation is now compact under the condition that \(G_{\rm WW}\left( \omega_{0} \right)\) is compact. This equation, or variants on it, then provide a powerful starting point to address this problem. For instance, consider the time-domain wave equation:

\begin{equation}
    \left( c^{- 2}\partial_{z}^{2} - \partial_{t}^{2} \right)E = \partial_{t}^{2}\chi E,
    \label{eq:SeparatedCavityWaveEquation}
\end{equation}

\noindent where \(E\) obeys the Floquet ansatz (see Equation (\ref{eq:FloquetAnsatz})) over a single period \(T\), with outgoing spatial boundary conditions. Following Equation (\ref{eq:BornEigenvalues}), we then consider applying \(\left( c^{- 2}\partial_{z}^{2} - \partial_{t}^{2} \right)^{- 1}\), where the inverse is obtained with the Green's function,

\begin{multline}
    G\left( z,t;\omega_{0} \right) = - \frac{1}{2c}\sum_{n = 0}^{\left\lceil \frac{L}{cT} \right\rceil}{\Theta\left( c(t + nT) - |z| \right){\rm e}^{\mathrm{i}n\omega_{0}T}} \\ + \frac{{\rm e}^{\mathrm{i}\left( \left\lceil \frac{L}{cT} \right\rceil + 1 \right)\omega_{0}T}}{1 - {\rm e}^{\mathrm{i}\omega_{0}T}},
    \label{eq:ExampleGreensFunction}
\end{multline}

\noindent which holds away from \({\rm e}^{\mathrm{i}\omega_{0}T} = 1\) and \(\omega_{0} = \infty\), corresponding directly to the spectrum of the wave equation, and where \(L\) is the length of a contiguous region containing the entire cavity. As a result, \(\left( c^{- 2}\partial_{z}^{2} - \partial_{t}^{2} \right)^{- 1}\) can be expressed as an integral operator with a square integrable kernel, from which it follows that it is in fact a valid Hilbert-Schmidt integral operator\cite{Zweck2025}, and thus compact\cite{Hanson2002}. By applying the property that a compact operator multiplied by a bounded operator is still compact, it follows that the FQNM condition

\[\left( 1 - \left( c^{- 2}\partial_{z}^{2} - \partial_{t}^{2} \right)^{- 1}\partial_{t}^{2}\chi \right)E = 0\]

\noindent is compact as long as \(\partial_{t}^{2}\chi\) is bounded. Fortunately, this condition is particularly reasonable in physical materials where \(\chi\) corresponds to the response of massive, charged particles to an external field.

As a result, we cite the analytic Fredholm theorem~\cite{Renardy2004}, which proves that such generalised eigenvalues form a discrete set over the set of values where the operator \(A\left( \omega_{0} \right)\) is holomorphic. In our case, this corresponds to frequencies \(S_{0} = \mathbb{C}_{\infty} \setminus S_{\rm R}\), where \(S_{\rm R}\) contains the poles of \(G_{\rm WW}{\widetilde{L}}_{\rm WM}G_{\rm MM}{\widetilde{L}}_{\rm MW}\). In other words, accumulation points may occur only around the set of resonances of the decoupled system. Meanwhile, a theorem of Steinberg~\cite{Steinberg1968} proves that, as long as \(G_{\rm WW}{\widetilde{L}}_{\rm WM}G_{\rm MM}{\widetilde{L}}_{\rm MW}\) varies continuously, the solutions of Equation (\ref{eq:BornEigenvalues}) also vary continuously within \(S_{0}\), with any new modes appearing/disappearing at the edges of this set (i.e., at the decoupled resonances \(S_{\rm R}\)).

\end{document}